\documentclass[11pt]{article}
\usepackage[utf8]{inputenc}
\usepackage{graphicx}
\usepackage{xcolor}
\graphicspath{ {figs/} }
\usepackage{titlesec}
\usepackage[a4paper, total={6in, 9in}]{geometry}
\usepackage{float}
\usepackage{url}
\usepackage{hyperref}
\usepackage[scaled]{beramono}
\usepackage{dingbat}

\usepackage[scaled]{helvet}
 
\usepackage[T1]{fontenc}

\DeclareUnicodeCharacter{03B2}{beta}
\title{
{Engineering Supercomputing Platforms for Biomolecular Applications}
}
\author{Robert Welch, Charles Laughton, Oliver Henrich, Tom Burnley, Daniel Cole, Alan Real,\\ Sarah Harris, James Gebbie-Rayet}
\date{}

\begin{document}

\def\arraystretch{1.25}

\renewcommand{\contentsname}{\vspace{-32pt}}

\maketitle

\tableofcontents
\newpage

\section{Executive Summary}

The computational biology community utilises a diverse range of methods that all have different computational requirements, including CPU, GPU, storage and memory. A diverse range of hardware is needed to meet all of these needs. Diverse hardware also makes HPC more resilient against change, both external (e.g. increasing hardware costs, vendor lock-in) and internal (e.g. adoption of new methods).
    \begin{itemize}
        \item AMD's Instinct GPUs cost less than Nvidia's offerings, and their software is supported by most computational biology methods. However, this software is significantly less mature than Nvidia's, and supporting it would require more time and energy from software developers and system administrators to compensate.
        \item Specialised AI hardware, such as Nvidia's Grace Hopper `superchips', are actually general-purpose computers and offer excellent performance in most computational biology methods. In computational biology, AI hardware is typically used in tandem with more traditional computational biology methods, as part of a larger workflow, so these machines can be offered to users for a broad range of applications, and not solely AI.
        \item Increasing energy costs and environmental concerns mean that efficient HPC usage is just as important as raw speed. GPUs offer the most efficient way to run most computational biology tasks, though some tasks still require CPUs. For users, confining jobs to a single HPC node or GPU where possible is the best way to maximise efficiency.
        \item A contemporary HPC node running molecular dynamics can produce around 10GB of data per day. Investments in data infrasturcture by the EPSRC and UKRI should be embedded into the comunities focused on HPC. Infrastructure to transfer this data around will become increasingly important.
    \end{itemize}

As the HPC landscape has become more complex, it has become more difficult and labour-intensive for system administrators to deploy software and keep systems online, leading to high downtime and a paucity of available software. 
    \begin{itemize}
    \item An increased focus on DevOps, including recruiting new talent and training/upskilling,  and increased collaboration and knowledge-sharing between HPC centres could increase HPC uptime and reduce duplicated work between HPC centres.
    \item The consortium model for HPC communities has been extremely successful in user training and in providing specialised support to HPC admins, and expanding it would ease the burden on system administrators and allow communities that use HPC to be supported directly by people from those communities.
    \item Build frameworks and virtualisation/containerisation tools such as EasyBuild and podman-hpc could further ease the burden on system administrators, allowing users to easily set up their own software environment instead of relying on pre-installed modules.
    \item Some computational biology communities (e.g. cryo-EM) are poorly-supported or unsupported on HPC, even though they would greatly benefit from these resources.
\end{itemize}
\newpage

\section{Introduction}\label{sec:intro}

Computer simulations allow us to understand how biological molecules work --- how drugs bind to their biological targets, how enzymes catalyse reactions, and how proteins fold into their functional forms~\cite{huggins_biomolecular_2019,mey_best_2020,lonsdale_practical_2012,lindorff-larsen_how_2011}. This knowledge is foundational to fields such as biotechnology, drug development and cancer research. Running these simulations requires powerful supercomputers \cite{mulholland_large-scale_2024}, and without adequate investment in supercomputing, the UK risks falling behind in this key research area.\\

In the past, staying ahead meant one thing: building a bigger, faster machine, which was traditionally represented by a performance metric called `floating point operations per second', or FLOPS. As computers have become more powerful, FLOPS has become a less useful metric than it once was. Supercomputers are complicated and heterogeneous, and implement many different architectures and designs. The exact performance of these machines depends on a wide range of factors: the software being used, the architecture of the machine, and user configuration.\\

The computational biology community has very diverse computational needs, and a supercomputer with the wrong hardware, or improperly configured software will not serve those needs, irrespective of its raw performance. This report will outline those needs, and show how well they are met by different hardware and software configurations. Section \ref{sec:hpcsystems} will outline the test simulations and machines that were used. Sections \ref{sec:MD}, \ref{sec:chemistry} and \ref{sec:EM} show the performance for a representative selection of computational biology methods and supercomputer hardware. Finally, section \ref{sec:software} will discuss ways to make supercomputers easier to use and manage. The aim of this report is to provide this information as a reference for users, system administrators and system builders. In addition to the executive summary above, the results will be discussed in more detail in section \ref{sec:conclusion}.\\

\subsection{HPC Systems Tested}\label{sec:hpcsystems}

Four High Performance Computing systems (HPC systems) were used for benchmarking. These benchmarks do not feature direct, controlled comparisons between specific pieces of hardware, instead they aim to represent the widest possible range of HPC hardware. The specifications of each machine are summarised in table \ref{tab:systems}.

\begin{table}[h]
\centering
\begin{tabular}{lllll}
         & \href{https://www.archer2.ac.uk/}{ARCHER2}      & \href{https://www.jade.ac.uk/}{JADE2}         & \href{https://n8cir.org.uk/bede/}{BEDE-GH}      & \href{https://lumi-supercomputer.eu/}{LUMI-G}      \\
\hline
CPU      & 2x EPYC 7742 & 2x E5-2698    & ARM Neoverse & EPYC 7A53   \\
GPU      & n/a          & 8x Tesla V100 & H100         & 4x MI250x      \\
RAM      & 256/512gb    & 512gb         & 480gb        & 512gb   \\
Platform & HPE Cray EX  & DGX MAX-Q     & n/a          & HPE Cray EX       
\end{tabular}
    \caption{HPC systems tested}
    \label{tab:systems}
\end{table}

ARCHER2 and JADE2 are both older facilities, and are tested as `reference' CPU and GPU systems, which newer systems can be compared to. The BEDE-GH cluster is an Nvidia Grace Hopper (GH200) testbed system offered by BEDE. LUMI-G was used to test AMD's hardware and software, and while LUMI-G nodes feature the older MI250X GPUs (meaning they can't be directly compared to the GH200), the limiting factor with AMD hardware is more likely to be software compatibility than performance, which is discussed in Section \ref{sec:software}. In addition to the systems mentioned in table \ref{tab:systems}, some supplemental tests were run on the University of Nottingham `Ada' cluster and the GH200 testbed at ISAMBARD-AI.\\

\section{Molecular Dynamics Benchmarks}\label{sec:MD}

\begin{figure}[H]
    \centering
    \includegraphics[width=\textwidth]{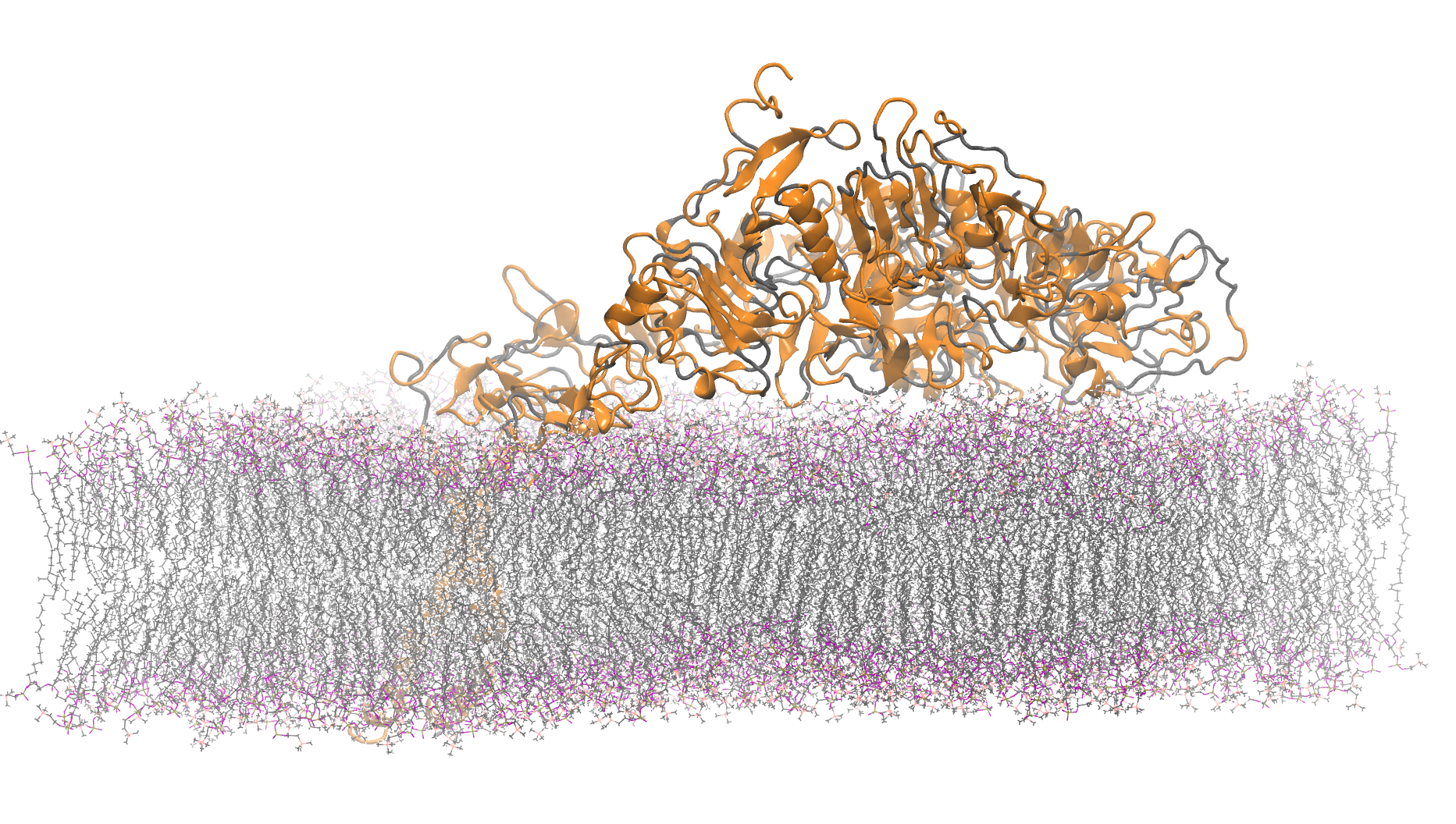}
    \caption{Example rendering of a molecular dynamics system featuring protein and lipid (hEGFR Dimer of 1IVO and 1NQL described in table \ref{tab:hecbms}).}
    \label{fig:mdsample}
\end{figure}

According to ARCHER2 user data \cite{andy_turner_software_2022}, Molecular Dynamics (MD) is the most common application for computational biology on HPC, and the third most common application overall, behind materials science and fluid dynamics. MD simulations are typically composed of biomolecules (proteins, lipids, and/or DNA) suspended in solvent (usually water). These molecules are represented down to the level of atoms and interatomic bonds, whose interactions are parameterised by classical `force fields' --- lookup tables for interaction energies, taken from quantum chemistry calculations. Calculating the forces that act on these atoms, from interatomic bonds and electric charge, accounts for most of the computational work in MD simulations \cite{andersson_breaking_2023}. Atomistic MD typically takes place on length scales of hundreds of nanometers and and time scales of nanoseconds to microseconds, and its performance is usually measured in ns/day: the number of nanoseconds of simulation time that can be achieved in one day of real time.

\subsection{Comparison of MD Software}

Molecular Dynamics software is highly consolidated --- almost every MD user is using one of five software packages: GROMACS \cite{abraham_gromacs_2015}, AMBER \cite{case_ambertools_2023}, NAMD \cite{phillips_scalable_2020}, LAMMPS \cite{thompson_lammps_2022} and OpenMM \cite{eastman_openmm_2024}. These packages are broadly interoperable, though they all offer different features, implementing different force fields, simulation methods, interactions, constraint algorithms, and temperature/pressure controls \cite{poghosyan_gromacs_2006}. Because the software being tested implements different physics (and often yields different results) \cite{shirts_lessons_2016, eastman_openmm_2017, lee_charmm-gui_2015}, these benchmarks are not direct comparisons, though they are as close to one another as possible. More importantly, although this software is interoperable, it is not interchangeable --- it is not possible to easily substitute one program for another without changing the results. Therefore, it is not always useful to think of this comparison as \textit{`which molecular dynamics software is the fastest?'} because there are good reasons to use all of them. Instead, we want to see how well different HPC hardware supports MD software in aggregate.\\

The benchmarks in this section were run using the HecBioSim benchmark suite \cite{hecbiosim_hecbiosim_nodate}, a set of five MD simulations which are designed to be representative of common MD tasks. These systems are summarised in table \ref{tab:hecbms}.\\

\begin{table}[H]
\centering
\begin{tabular}{lllll}
System (PDB codes)                               & Atoms & Protein & Lipid & Water \\
\hline
3NIR Crambin \cite{schmidt_crystal_2011}                         & 21k         & 642           & 0           & 19k         \\
1WDN Glutamine-Binding Protein \cite{sun_structure_1998}        & 61k         & 3.5k          & 0           & 58k         \\
hEGFR Dimer of 1IVO and 1NQL \cite{ferguson_egf_2003, ogiso_crystal_2002} (fig \ref{fig:mdsample})         & 465k        & 22k           & 134k        & 309k        \\
Two hEGFR Dimers of 1IVO and 1NQL \cite{ferguson_egf_2003, ogiso_crystal_2002}    & 1.4M        & 43k           & 235k        & 1.1M        \\
Two hEGFR tetramers of 1IVO and 1NQL \cite{ferguson_egf_2003, ogiso_crystal_2002} & 3M          & 87k           & 868k        & 2M         
\end{tabular}
    \caption{HecBioSim Benchmark Suite. Total system sizes range from 21k to 3M atoms, and the chemistry is representative of that typically found in biological systems.}
    \label{tab:hecbms}
\end{table}

More information about the benchmark suite, MD configuration, system parameters, and compilation process can be found in the appendix (section \ref{sec:methodology}). A correction to the AMBER and OpenMM benchmark input files can be found in the corrigendum (section \ref{sec:corrigendum}).\\

\begin{figure}[H]
    \centering
    \includegraphics[width=\textwidth]{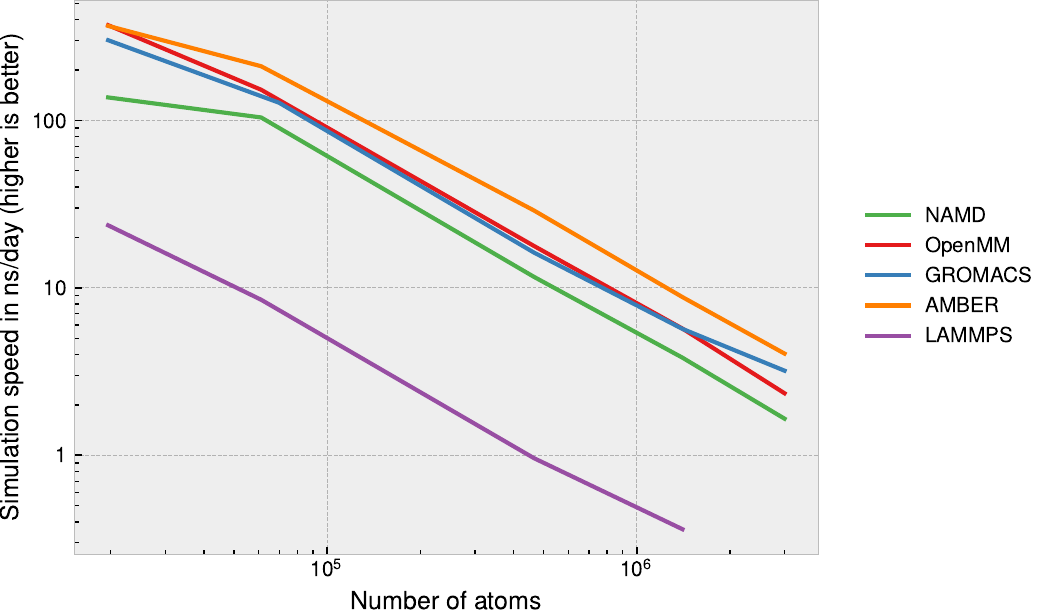}
    \caption{Molecular dynamics performance on a single GPU of JADE2 (higher ns/day is better).}
    \label{fig:jadeperformance}
\end{figure}

The benchmarks were first run on JADE2, a GPU machine with eight Nvidia V100 GPUs per node. Figure \ref{fig:jadeperformance} shows the performance on JADE2 using one-eighth of one node (one GPU and eight of the machine's 64 CPU cores).

All of the GPU benchmarks in this report were performed on single GPUs. Single GPUs are still the best way to run most MD jobs. On a technical level, not all MD software supports GPU domain decomposition, and those that do only see favourable scaling for extremely large (>10M atom) systems, which are not representative of typical MD jobs \cite{noauthor_massively_2023, noauthor_multi-gpu_nodate, noauthor_amber22_nodate}. On a philosophical level, molecular dynamics workloads tend not to be monolithic --- they are more often comprised of many small jobs, and many repeats of those jobs, rather than one big simulation. Breaking MD workloads into many parallel tasks is more efficient than scaling a small number of jobs up to many nodes \cite{andersson_breaking_2023}. Finally, on a practical level, JADE2's NVLink solution was broken at time of benchmarking.\\

\begin{figure}[H]
    \centering
    \includegraphics[width=\textwidth]{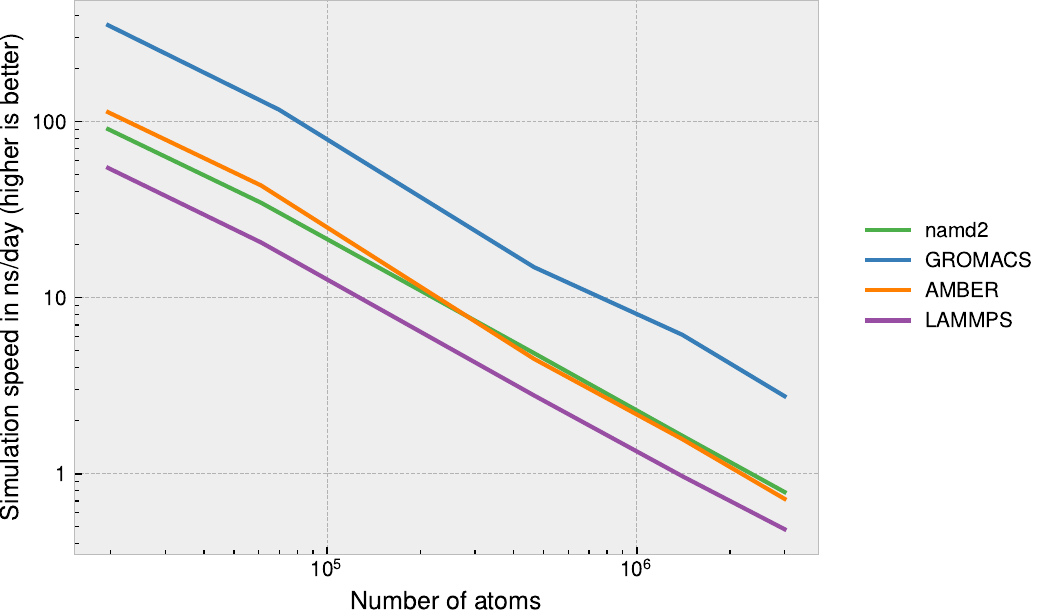}
    \caption{Molecular dynamics performance on a single node of ARCHER2 (higher ns/day is better).}
    \label{fig:archer2performance}
\end{figure}

\begin{figure}[H]
    \centering
    \includegraphics[width=\textwidth]{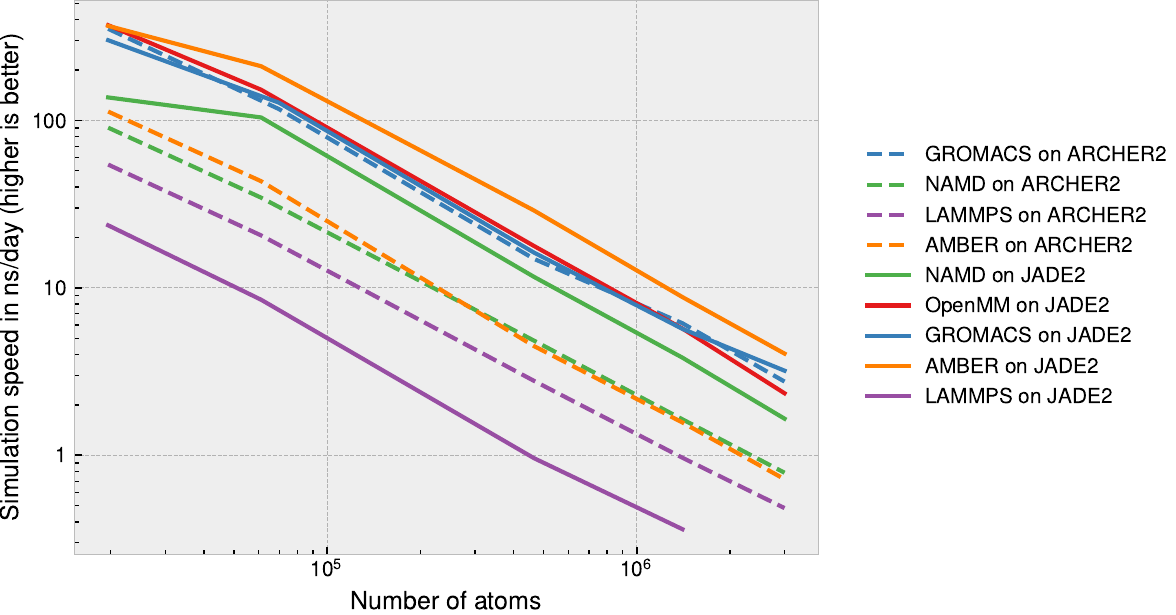}
    \caption{MD Performance comparison between a single node of ARCHER2 and single GPU on JADE2 (higher ns/day is better).}
    \label{fig:archer2jade2performance}
\end{figure}

The benchmark suite was also run on a single node of ARCHER2, which features 2 AMD EPYC 7742 64-core CPUs. The results are shown in figure \ref{fig:archer2performance}. Figure \ref{fig:archer2jade2performance} compares the performance between a single GPU of JADE2 and a single node of ARCHER2.\\

The overall performance of one node on ARCHER2 is similar to that of one GPU on JADE2. The most important result is to identify gaps in the performance for different applications --- in this case, LAMMPS is not optimised to run on GPUs, while  OpenMM\footnote{Not pictured, as OpenMM only supports a reference implementation on CPUs} is not optimised to run on CPUs.\\

\begin{figure}[H]
    \centering
    \includegraphics[width=\textwidth]{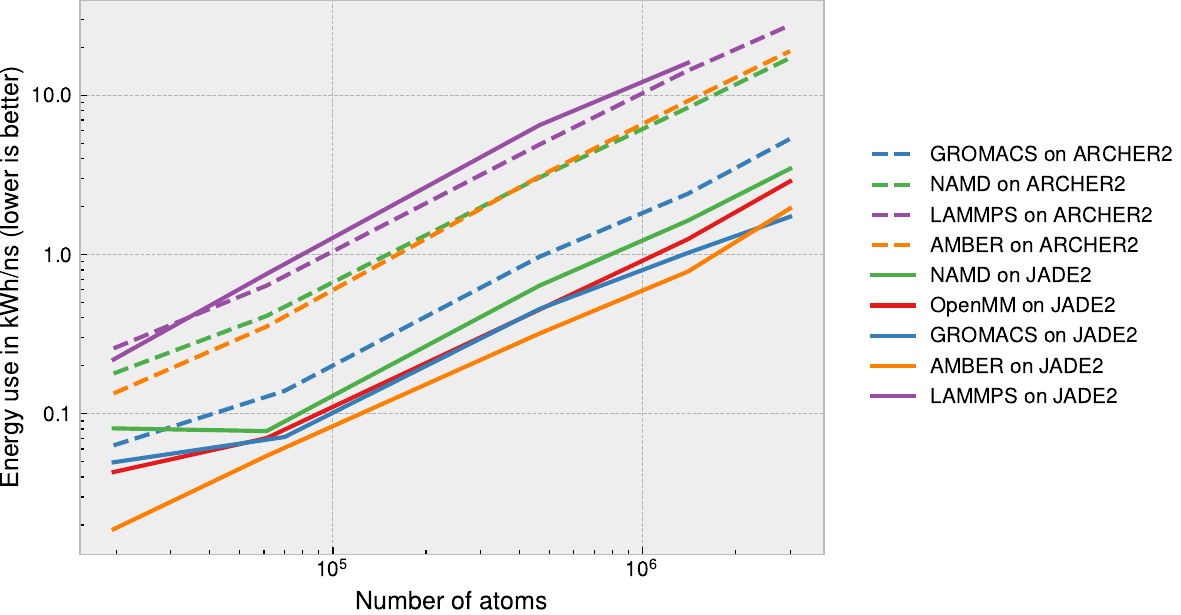}
    \caption{Comparison of MD energy use between a single node of ARCHER2 and a single node of JADE2 (Lower J/ns is more efficient).}
    \label{fig:archer2jade2efficiency}
\end{figure}

Figure \ref{fig:archer2jade2efficiency} shows the efficiency of ARCHER2 and JADE2 in J/ns, the total energy consumed by the job per nanosecond of simulation time. Although the performance of ARCHER2 and JADE are similar, JADE2 is more power efficient.\\

\subsection{MPI and OpenMP}

MPI and OpenMP are two protocols that allow for parallel computing: OpenMP allows for `shared memory' parallelisation, whereas MPI allows for `distributed memory' parallelisation. For a simulation running on multiple nodes, MPI can access memory on different nodes, whereas OpenMP can only access memory within a single node. However, the extra complexity of the MPI protocol can incur more overhead than OpenMP.\\

GROMACS, NAMD and LAMMPS all implement mixed OpenMP and MPI parallelisation. To find the optimum balance between MPI and OpenMP, benchmarks were run using different combinations of MPI processes and OpenMP threads, for between 1 and 16 nodes. The results can be found in section \ref{sec:archer2ompappendix}. In almost all cases, the benchmarks ran best using as many MPI processes as there are available CPU cores, and a single OpenMP thread per process. Multiple threads performed best in some edge cases, such as when a small system is spread over many nodes.\\

\begin{figure}[H]
    \centering
    \includegraphics[width=\textwidth]{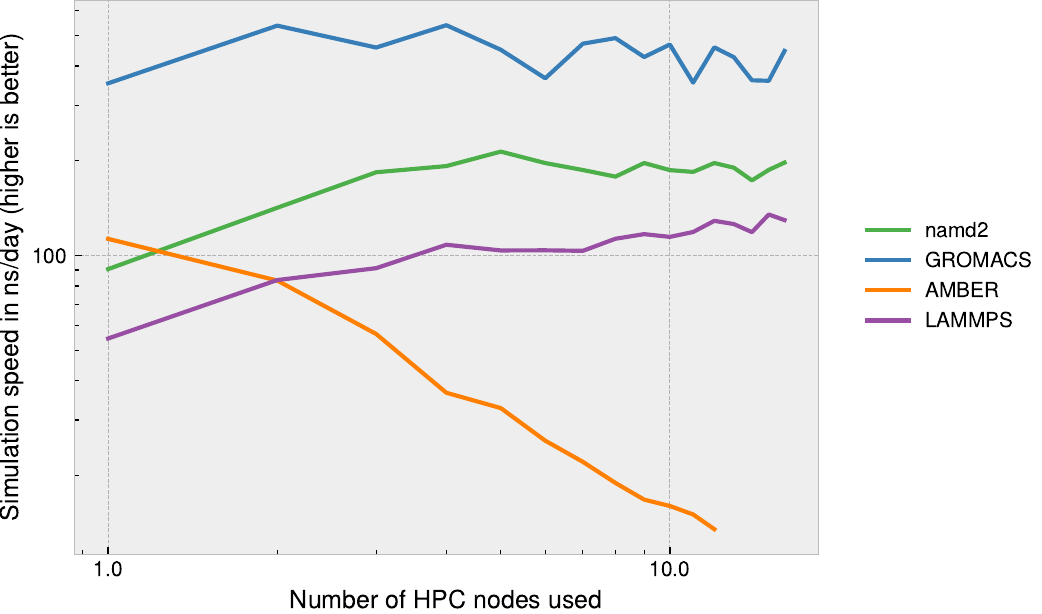}
    \caption{Performance of different MD software on ARCHER2 for the 20k atom system, running on up to 16 nodes. Higher ns/day is better.}
    \label{fig:archer2nodes20k}
\end{figure}

\begin{figure}[H]
    \centering
    \includegraphics[width=\textwidth]{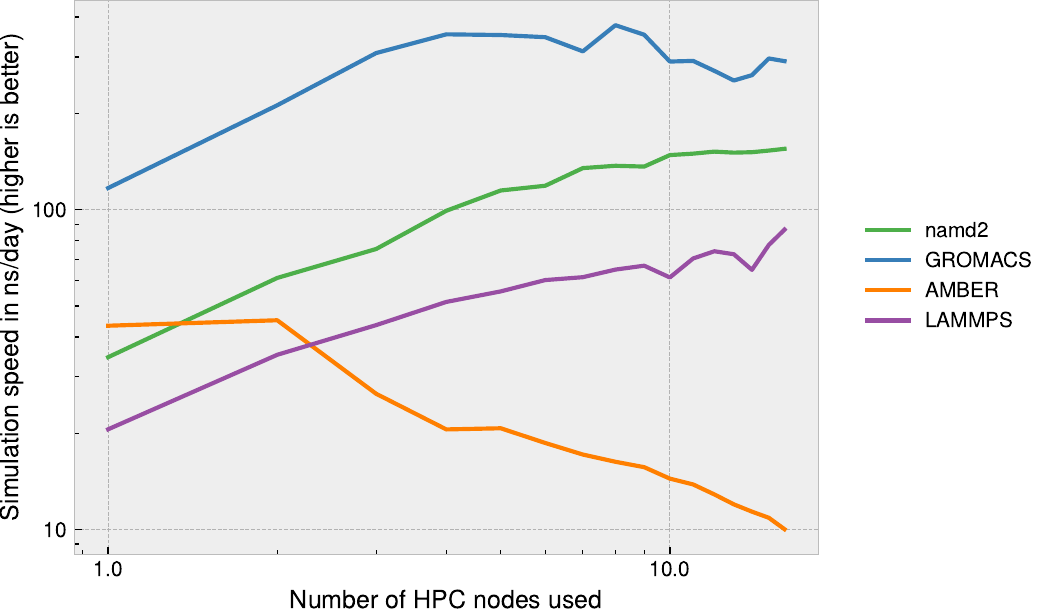}
    \caption{Performance of different MD software on ARCHER2 for the 61k atom system, running on up to 16 nodes. Higher ns/day is better.}
    \label{fig:archer2nodes61k}
\end{figure}

\begin{figure}[H]
    \centering
    \includegraphics[width=\textwidth]{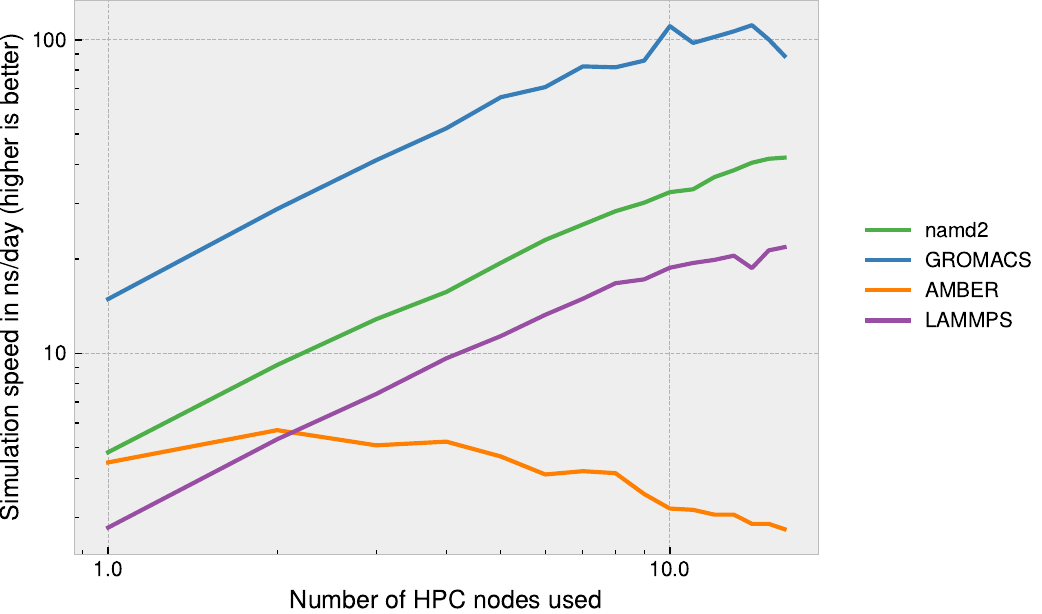}
    \caption{Performance of different MD software on ARCHER2 for the 465k atom system, running on up to 16 nodes. Higher ns/day is better.}
    \label{fig:archer2nodes465k}
\end{figure}

\begin{figure}[H]
    \centering
    \includegraphics[width=\textwidth]{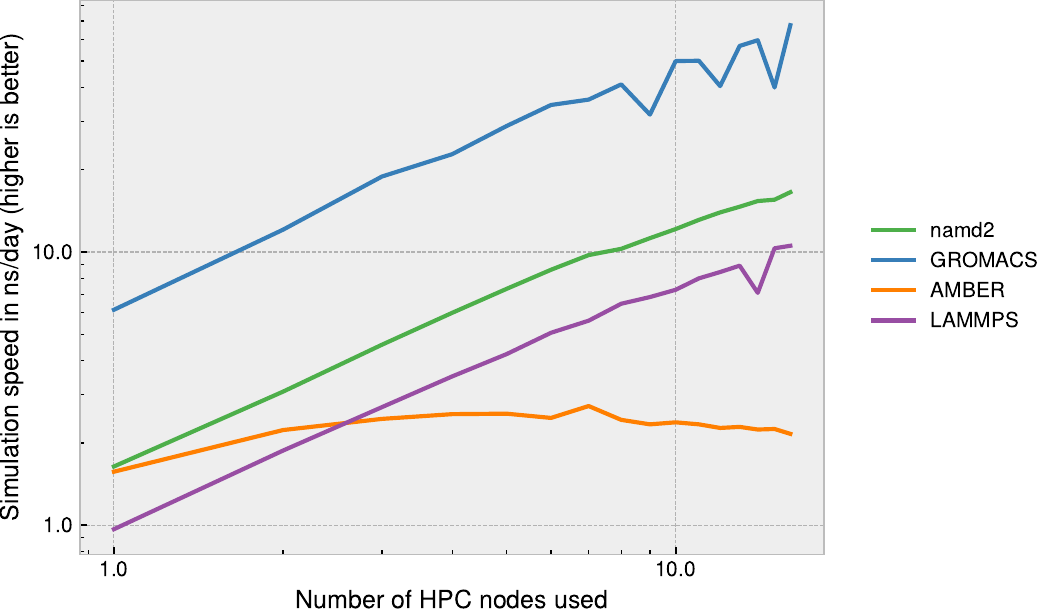}
    \caption{Performance of different MD software on ARCHER2 for the 1400k atom system, running on up to 16 nodes. Higher ns/day is better.}
    \label{fig:archer2nodes1400k}
\end{figure}

\begin{figure}[H]
    \centering
    \includegraphics[width=\textwidth]{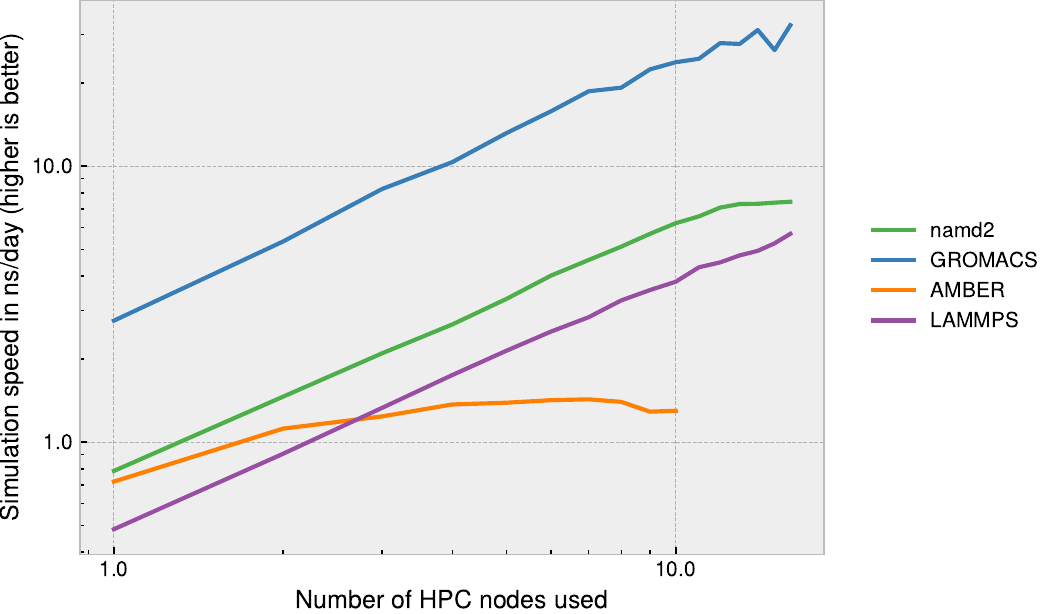}
    \caption{Performance of different MD software on ARCHER2 for the 3000k atom system, running on up to 16 nodes. Higher ns/day is better.}
    \label{fig:archer2nodes3000k}
\end{figure}

Figures \ref{fig:archer2nodes20k}-\ref{fig:archer2nodes3000k} show the `best case' performance (using the optimal combination of MPI and OpenMP) for each MD program, for up to sixteen nodes.\\

OpenMM and AMBER are both designed only to run on GPUs, so aren't performant on CPU system such as ARCHER2. GROMACS, NAMD and LAMMPS all have similar scaling properties --- the scaling for the 20k system is poor, the 61k system scales well up to around 8 nodes, and larger systems scale linearly up to 16 nodes. \\

\begin{figure}[H]
    \centering
    \includegraphics[width=\textwidth]{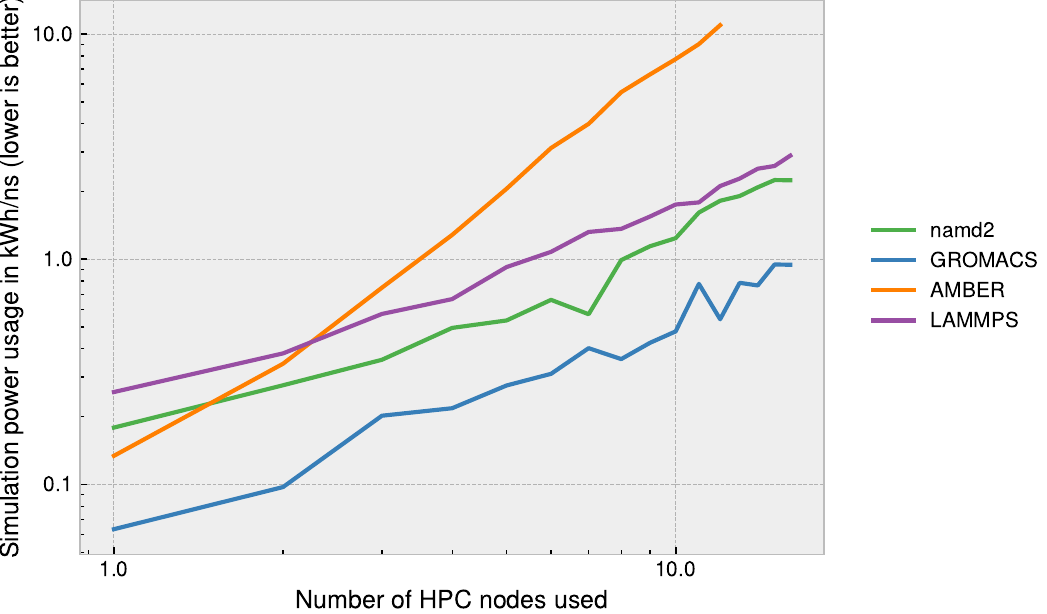}
    \caption{Energy use of different MD software on ARCHER2 for the 20k atom system, running on up to 16 nodes. Lower kWh/ns is better.}
    \label{fig:archer2energy20k}
\end{figure}

\begin{figure}[H]
    \centering
    \includegraphics[width=\textwidth]{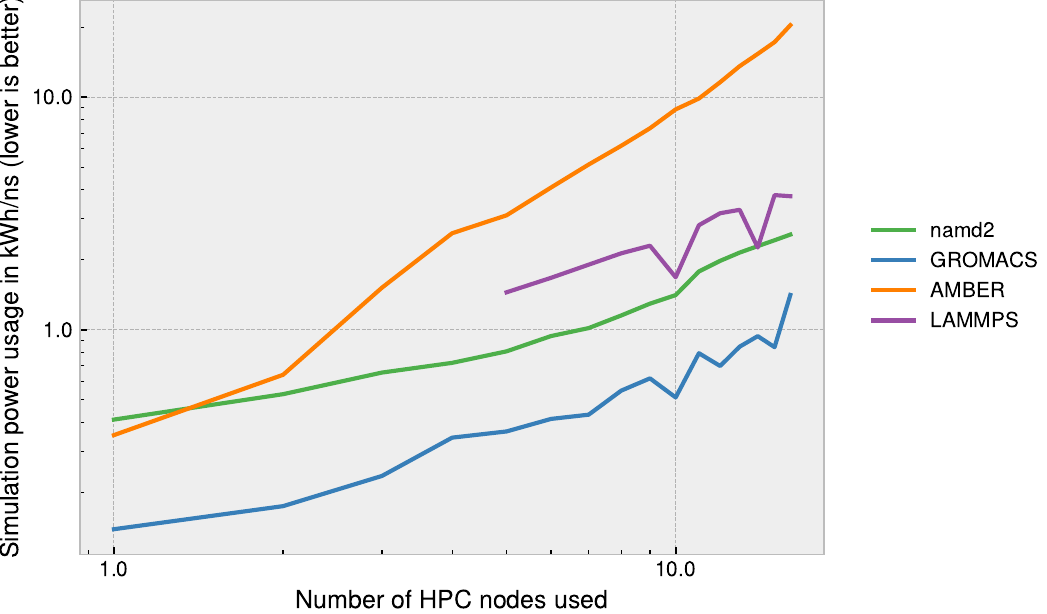}
    \caption{Energy use of different MD software on ARCHER2 for the 61k atom system, running on up to 16 nodes. Lower kWh/ns is better.}
    \label{fig:archer2energy61k}
\end{figure}

\begin{figure}[H]
    \centering
    \includegraphics[width=\textwidth]{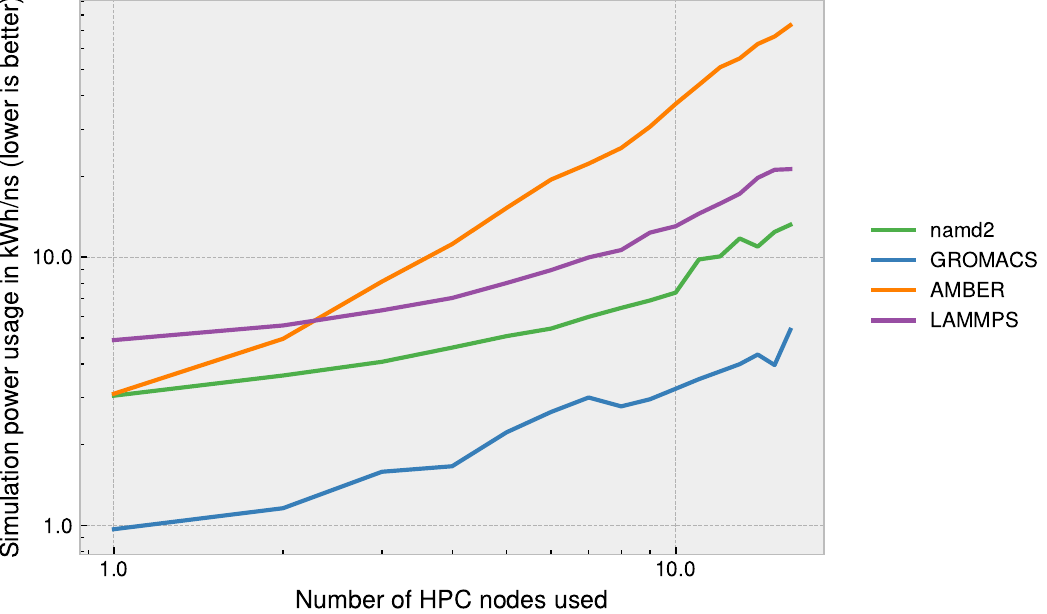}
    \caption{Energy use of different MD software on ARCHER2 for the 465k atom system, running on up to 16 nodes. Lower J/ns is better.}
    \label{fig:archer2energy465k}
\end{figure}

\begin{figure}[H]
    \centering
    \includegraphics[width=\textwidth]{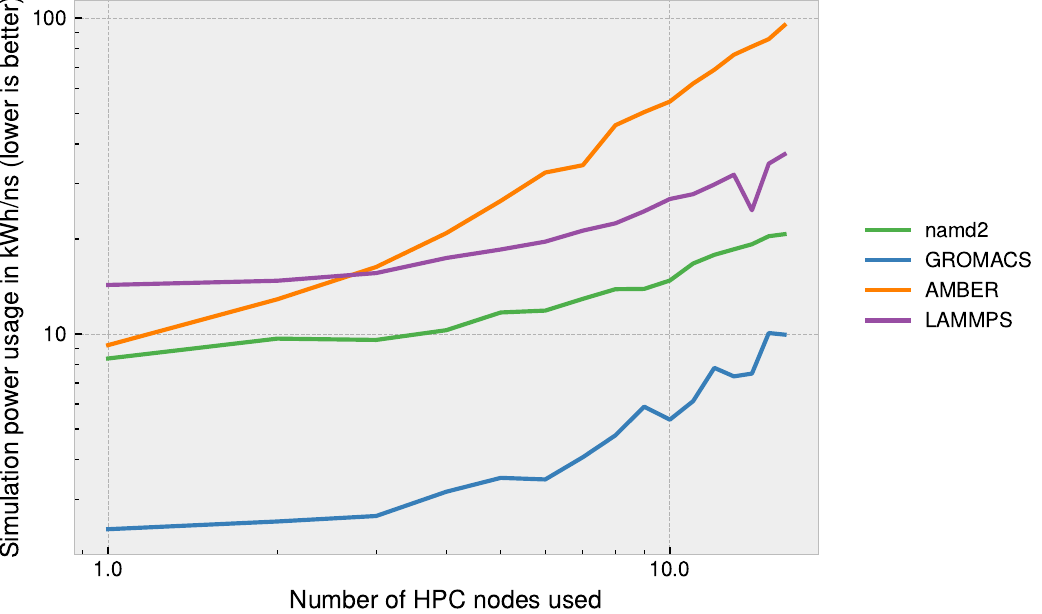}
    \caption{Energy use of different MD software on ARCHER2 for the 1400k atom system, running on up to 16 nodes. Lower kWh/ns is better.}
    \label{fig:archer2energy1400k}
\end{figure}

\begin{figure}[H]
    \centering
    \includegraphics[width=\textwidth]{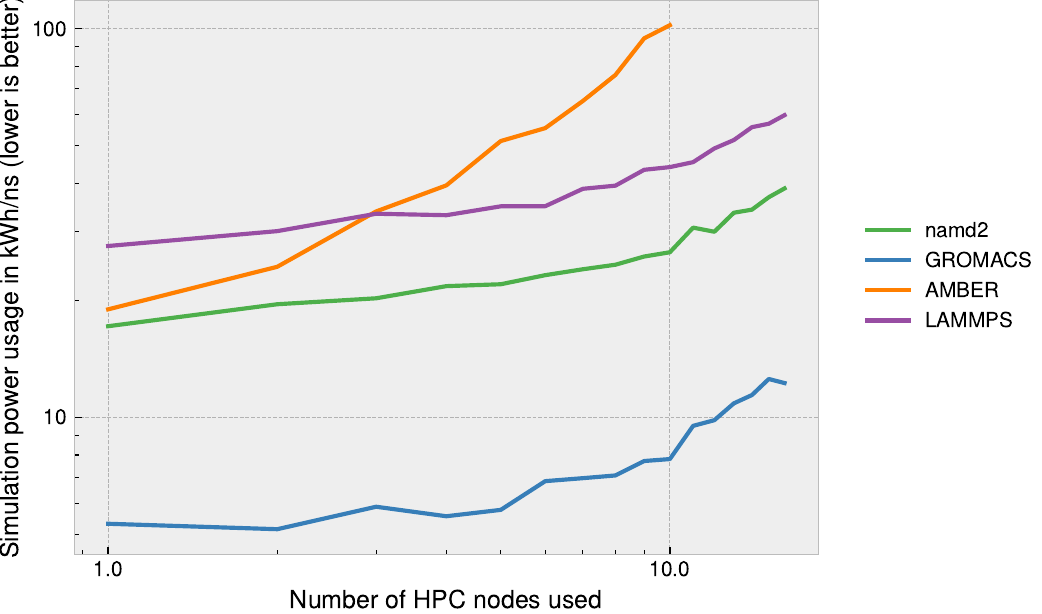}
    \caption{Energy use of different MD software on ARCHER2 for the 3000k atom system, running on up to 16 nodes. Lower kWh/ns is better.}
    \label{fig:archer2energy3000k}
\end{figure}

Scaling to many nodes still incurs a cost to performance (and efficiency) which is not adequately represented by plots showing only the scaling performance. Figures \ref{fig:archer2energy20k}-\ref{fig:archer2energy3000k} show how scaling to many nodes can impact the energy efficiency of the benchmark. Even with large systems, scaling to many nodes can triple the energy usage per nanosecond of simulation time.\\

\subsection{GH200}

Historically, the speed of computer memory has lagged behind the speed of CPUs \cite{mckee_reflections_2004}. Significant increases in performance have been driven by techniques such as instruction pipelineing and out-of-order execution, which aim to decrease the time spent waiting for data to be retrieved from RAM \cite{hager_introduction_2010}. The need for systems with fast, plentiful memory for AI applications has led to the development of platforms such as the Nvidia GH200 and AMD MI300X, which combine CPUs and GPUs onto a single board, with fast memory and a high-speed interconnect between the CPU and GPU memory. Though these units are marketed as `AI accelerators', they are actually general-purpose computers, and are extremely performant for many common computing tasks, including MD.\\

\begin{figure}[H]
    \centering
    \includegraphics[width=\textwidth]{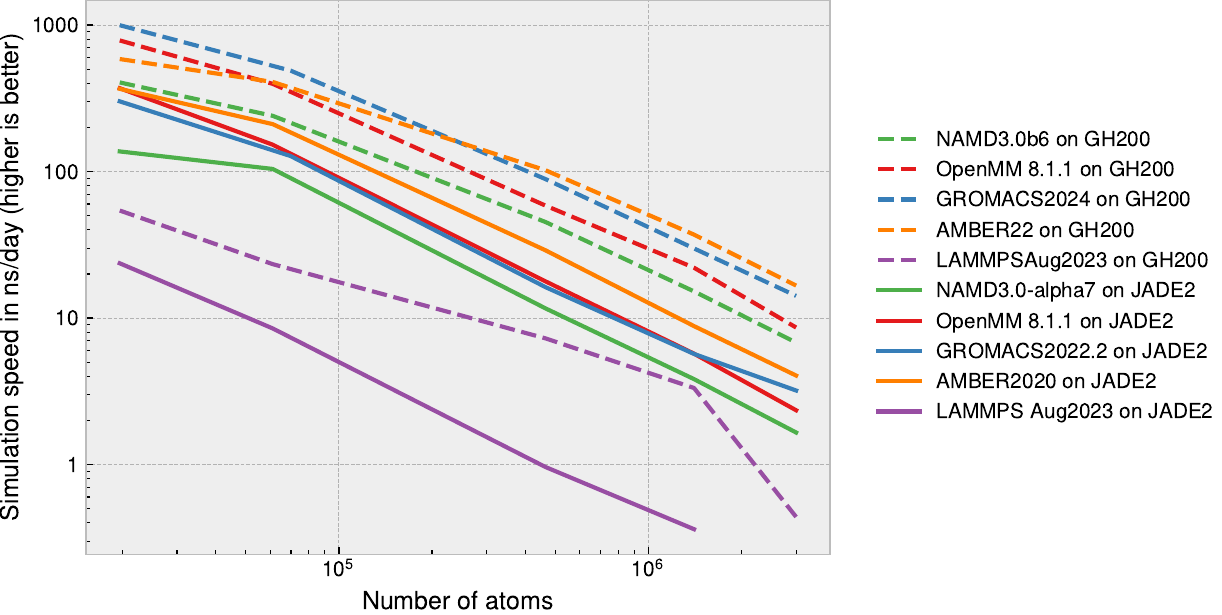}
    \caption{Performance comparison for different MD software between JADE2 (dashed lines) and the BEDE GH200 testbed (solid lines) (higher ns/day is better).}
    \label{fig:bedejade2performance}
\end{figure}

Nvidia's Grace Hopper `superchip' features 72 ARM Neoverse V2 cores and a `Hopper' series GPU. The BEDE-GH testbed features GH200 `superchips' with 480GB of CPU memory and 96GB of GPU memory. The five molecular dynamics programs listed in section \ref{sec:MD} were compiled on ARM and the benchmark suite listed in table \ref{tab:hecbms} was run. Figure \ref{fig:bedejade2performance} shows the results of the benchmarks on a single GH200 `superchip' compared to the older Nvidia V100 on JADE2.\\

The GH200 manages an order of magnitude better performance for every MD program, though the comparison is not exact, because the system libraries, compiler versions and software versions available on the GH200 machine are newer (see section \ref{sec:methodology} for more information on software versions).\\

\begin{figure}[H]
    \centering
    \includegraphics[width=\textwidth]{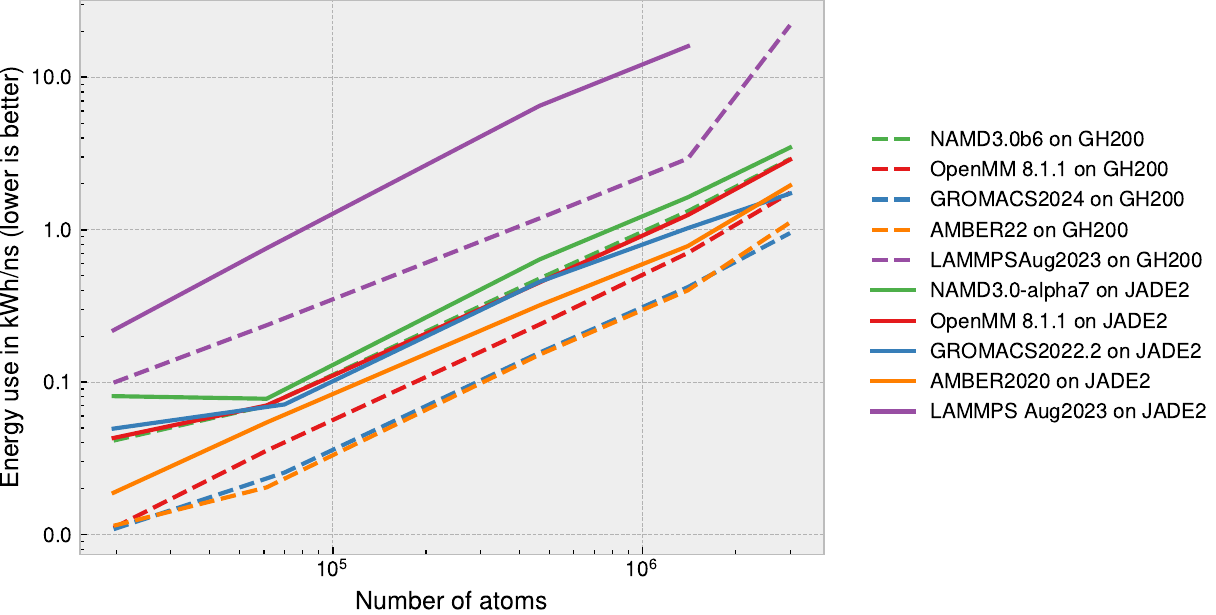}
    \caption{Energy use different MD software between JADE2 (dashed lines) and the BEDE GH200 testbed (solid lines) (lower kWh/ns is better).}
    \label{fig:bedejade2efficiency}
\end{figure}

Figure \ref{fig:bedejade2efficiency} shows the energy use of molecular dynamics on the GH200, compared to JADE2.\\

The HPC systems being compared in this report all have different power requirements. The power efficiency allows us to compare them in a way that accounts for the power usage (and environmental impact) of these machines. While the GH200 does use more power, it is still more efficient than its predecessors.\\

The rise in AI applications means that many datacenter GPUs are designed according to the needs of AI workloads, and so are over-specced for MD applications. Third-party benchmarks find that Grace Hopper's MD performance is similar to the performance of their (cheaper) discrete GPUs, such as the A100 \cite{noauthor_namd_nodate, technologies_molecular_2024, noauthor_amber_nodate, noauthor_nvidia_nodate, pro_amber_nodate}. Additionally, most MD simulations do not use close to the 40-96GB of GPU memory offered by these cards (see figure \ref{fig:mdmemory}). Nvidia's Multi-Instance GPU (MIG) feature may provide a way to use all of this memory, as it allows multiple simulations to run on the same GPU with minimal degradation in performance. This will be discussed in section \ref{sec:mig}.\\

\subsection{AMD MI250X}

\begin{figure}[H]
    \centering
    \includegraphics[width=\textwidth]{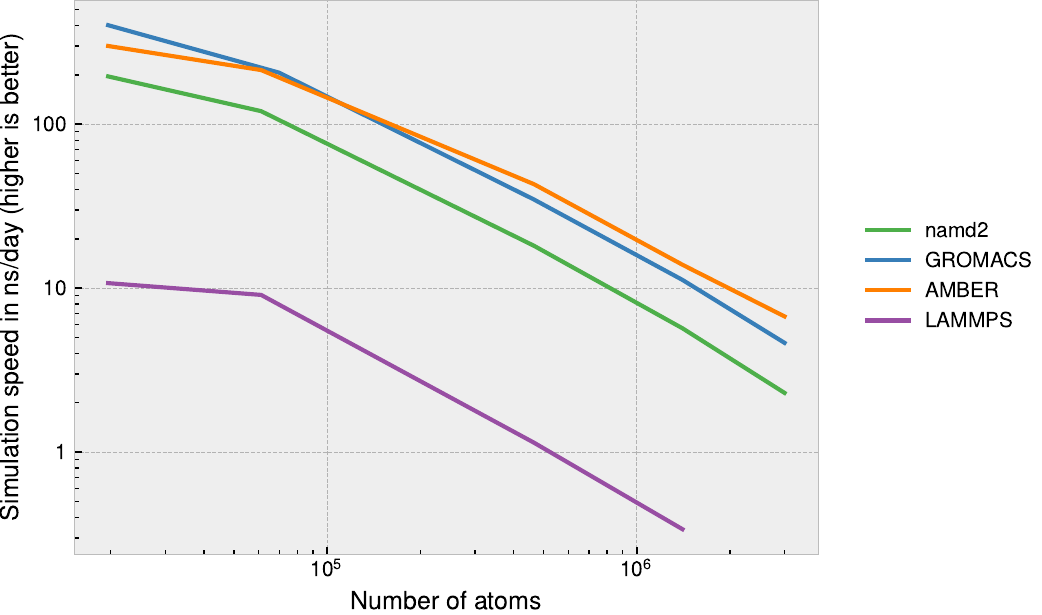}
    \caption{Molecular dynamics performance on a single GPU of LUMI-G (higher ns/day is better).}
    \label{fig:lumigperformance}
\end{figure}

AMD Instinct GPUs and accelerators offer raw performance comparable to their Nvidia counterparts, but cost significantly less. This makes them an appealing alternative to Nvidia, who have held a near-monopoly in this space for the last decade \cite{noauthor_november_nodate}. However, these GPUs are not a like-for-like alternative. Their ROCm software stack is less mature than Nvidia's CUDA and is still not fully supported by all MD software. Additionally, as will be discussed in section \ref{sec:software}, supporting software on ROCm can be more difficult than on CUDA. Unfortunately, no working MI300X testbeds were available for this benchmarking project, so the closest possible alternative, the MI250X nodes on LUMI-G, were used instead. The performance of different MD software on LUMI-G is shown in figure \ref{fig:lumigperformance}.\\

\begin{figure}[H]
    \centering
    \includegraphics[width=\textwidth]{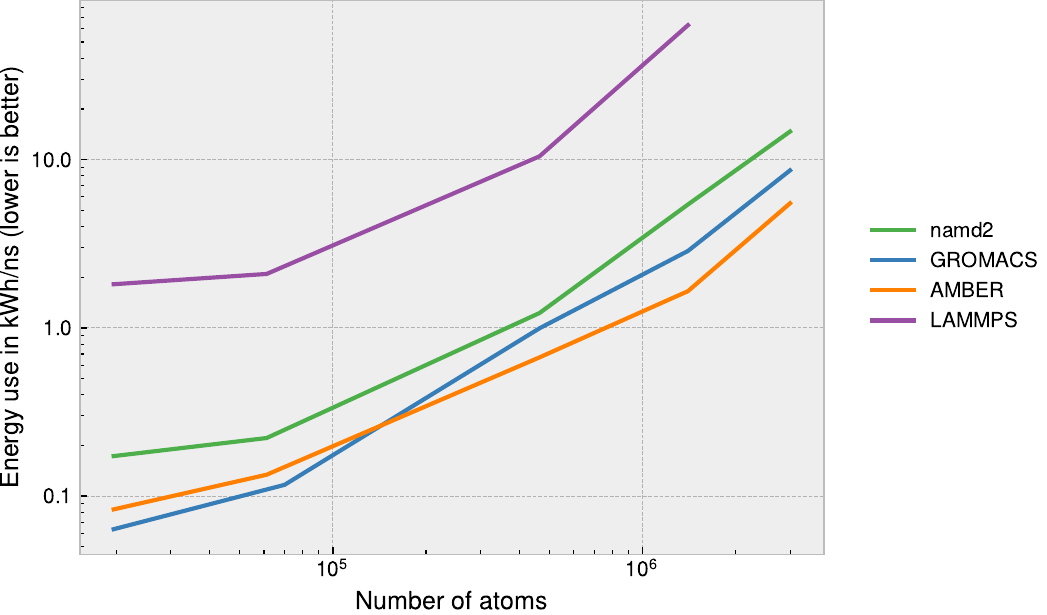}
    \caption{Molecular dynamics energy usage on a single GPU of LUMI-G (lower kWh/ns is better).}
    \label{fig:lumigefficiency}
\end{figure}

The MI250X's performance lies somewhere between JADE2 and the GH200, however, due to software compatibility issues, OpenMM could not be tested. Additionally, LAMMPS on LUMI-G is only supported via Kokkos, which does not support the exact combination of LAMMPS features used by the benchmark, so the LAMMPS benchmark on LUMI-G is not a direct comparison with LAMMPS on other systems. Figure \ref{fig:lumigefficiency} shows the efficiency of molecular dynamics on LUMI-G.\\

\subsection{Performance Across Different Hardware}

\begin{figure}[H]
    \centering
    \includegraphics[width=\textwidth]{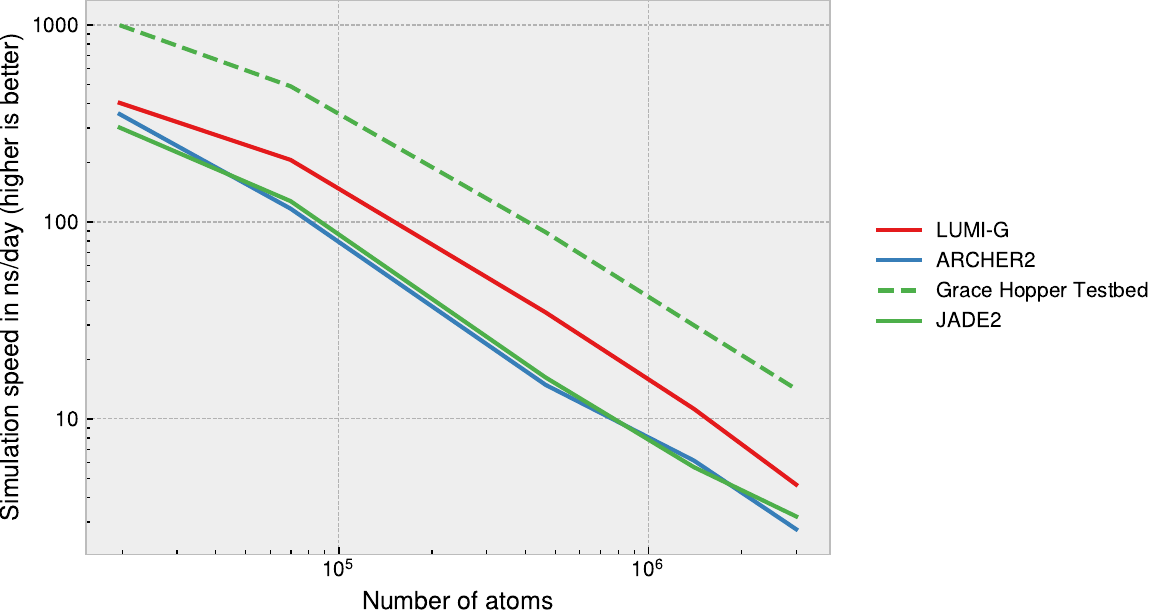}
    \caption{Comparison of GROMACS performance across multiple HPC systems (higher ns/day is better).}
    \label{fig:gromacsmachinecomparison}
\end{figure}

\begin{figure}[H]
    \centering
    \includegraphics[width=\textwidth]{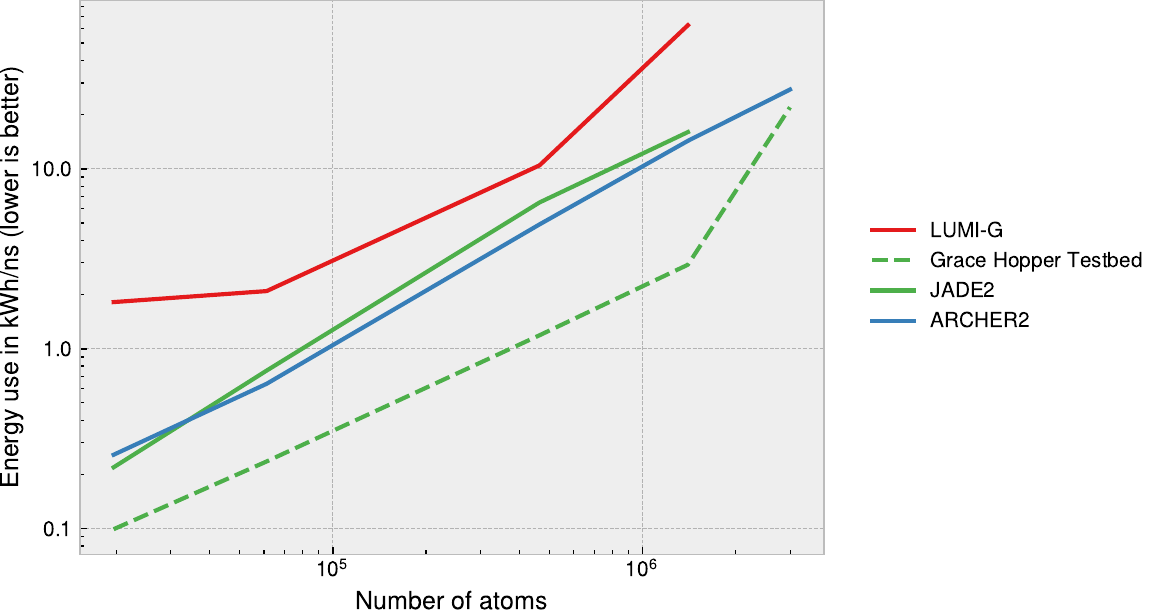}
    \caption{Comparison of LAMMPS energy usage across multiple HPC systems (lower kWh/ns is better).}
    \label{fig:lammpsmachineefficiency}
\end{figure}

For reference, figures \ref{fig:gromacsmachinecomparison} through to \ref{fig:lammpsmachineefficiency} show the performance and energy usage of MD software compared between different HPC systems.\\

\begin{figure}[H]
    \centering
    \includegraphics[width=\textwidth]{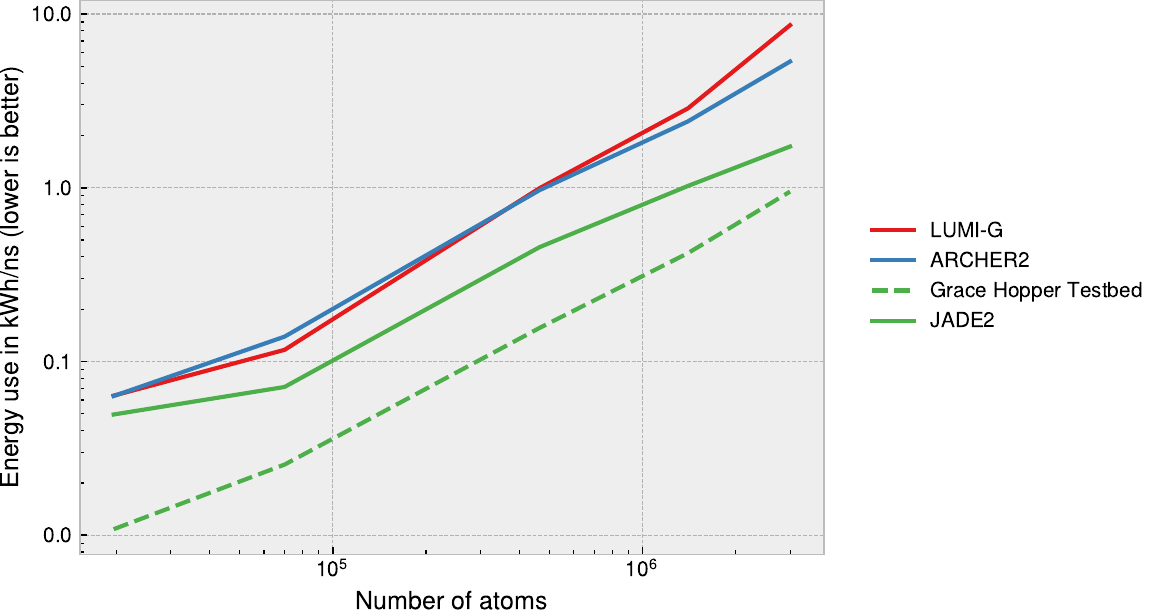}
    \caption{Comparison of GROMACS energy usage across multiple HPC systems (lower kWh/ns is better).}
    \label{fig:gromacsmachineefficiency}
\end{figure}

\begin{figure}[H]
    \centering
    \includegraphics[width=\textwidth]{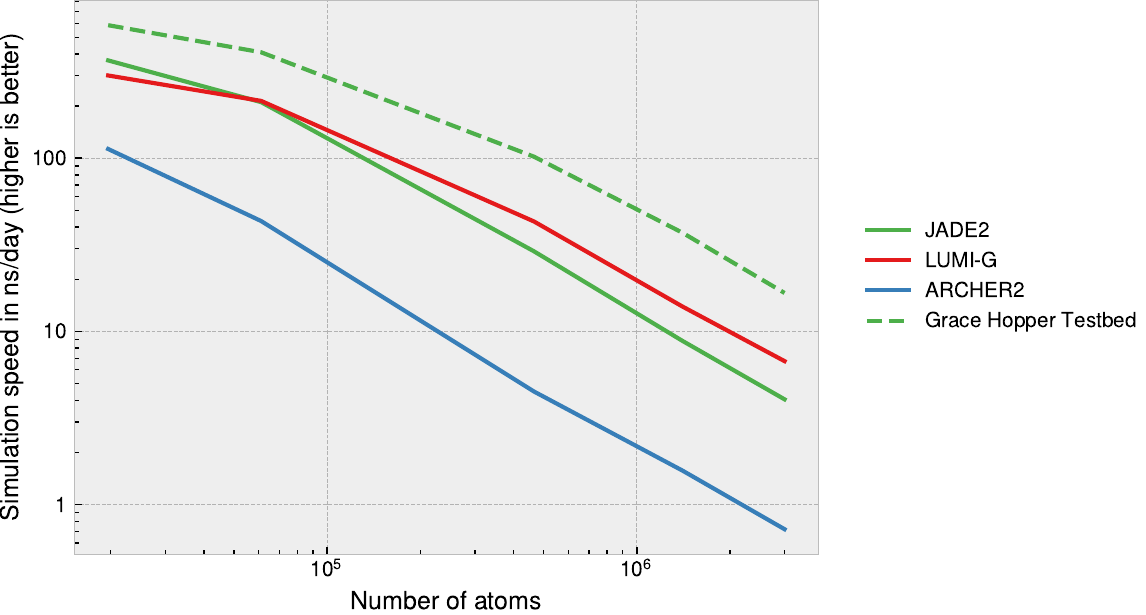}
    \caption{Comparison of AMBER performance across multiple HPC systems (higher ns/day is better).}
    \label{fig:ambermachinecomparison}
\end{figure}

\begin{figure}[H]
    \centering
    \includegraphics[width=\textwidth]{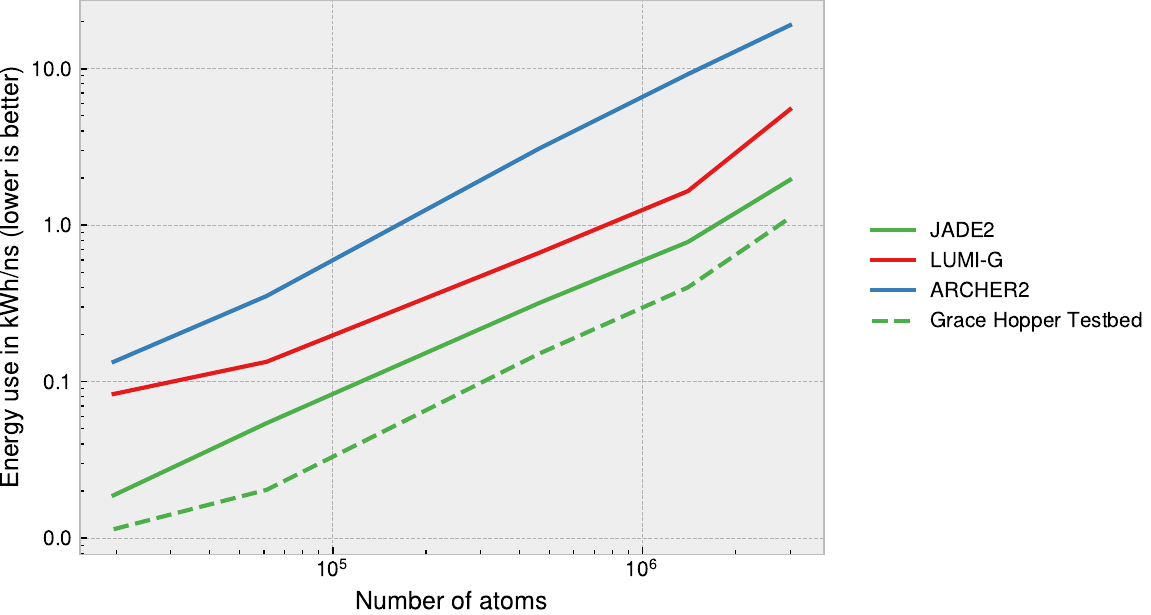}
    \caption{Comparison of AMBER energy usage across multiple HPC systems (lower kWh/ns is better).}
    \label{fig:ambermachineefficiency}
\end{figure}

\begin{figure}[H]
    \centering
    \includegraphics[width=\textwidth]{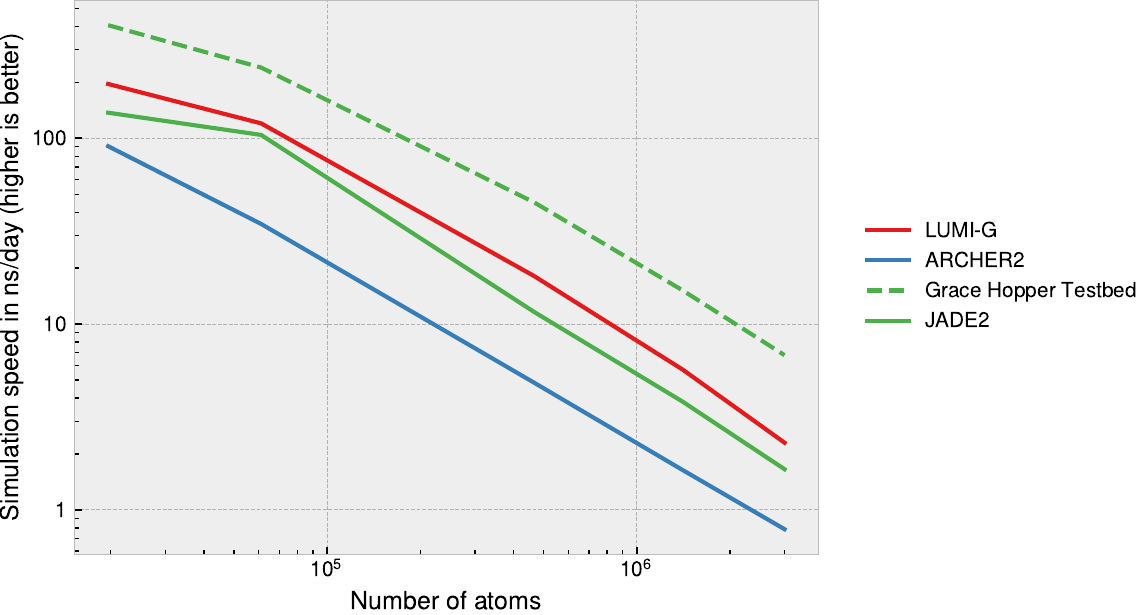}
    \caption{Comparison of NAMD performance across multiple HPC systems (higher ns/day is better).}
    \label{fig:namdmachinecomparison}
\end{figure}

\begin{figure}[H]
    \centering
    \includegraphics[width=\textwidth]{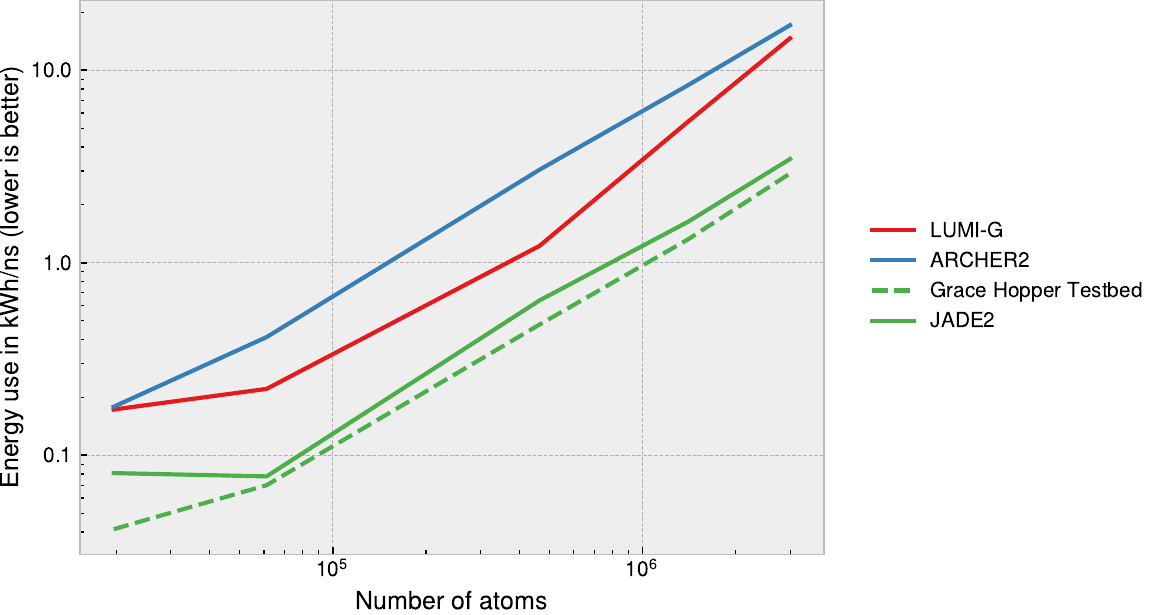}
    \caption{Comparison of NAMD energy usage across multiple HPC systems (lower kWh/ns is better).}
    \label{fig:namdmachineefficiency}
\end{figure}

\begin{figure}[H]
    \centering
    \includegraphics[width=\textwidth]{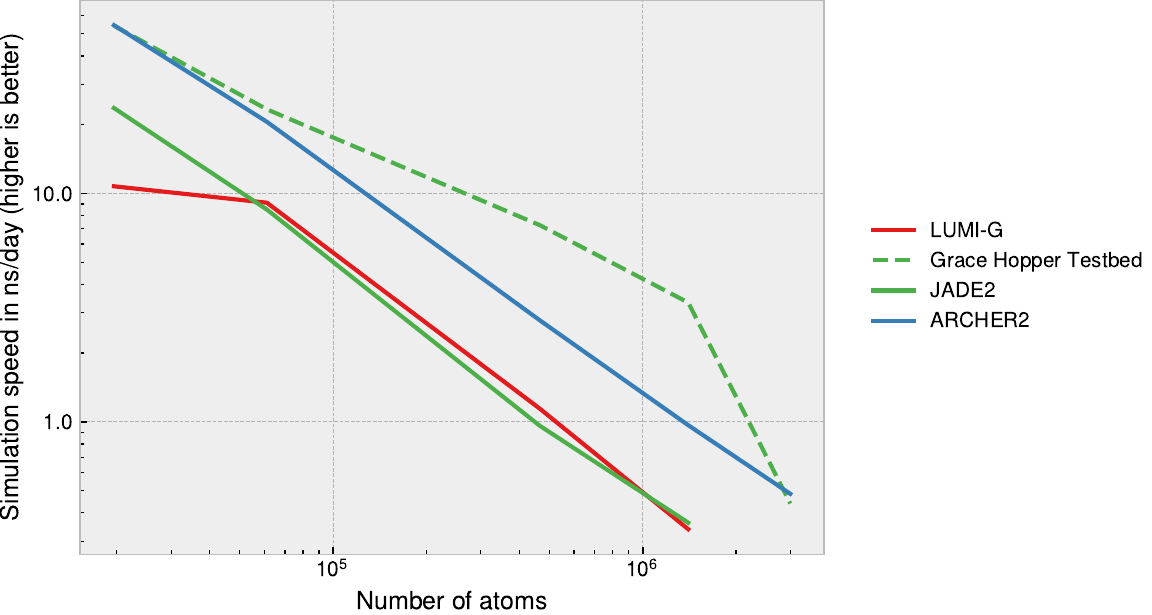}
    \caption{Comparison of LAMMPS performance across multiple HPC systems (higher ns/day is better).}
    \label{fig:lammpsmachinecomparison}
\end{figure}

In these benchmarks, we observe that LUMI-G is less efficient than other GPU-based systems, and sometimes CPU-based systems. The MI250X has a power draw significantly higher than its Nvidia counterpart, the A100, for roughly the same performance.\\

\subsection{Storage and Memory Requirements}\label{sec:storage}

Storage is an essential part of any HPC infrastructure, just as important as CPU and GPU compute. If storage arrays fill up, new HPC jobs can't run, and existing jobs risk failure or data corruption. The increase in computing power from new HPC systems will be accompanied by a proportional increase in the amount of storage space required. Table \ref{tab:storage} shows the storage required for a day's worth of MD simulations running on a single HPC node, for different levels of performance.\\

\begin{table}[h]
\centering
\begin{tabular}{llllll}
             & \begin{tabular}[c]{@{}l@{}}ns/day\\ per million atoms\end{tabular} & DCD   & HDF5  & NetCDF & \begin{tabular}[c]{@{}l@{}}XTC\\ (compressed)\end{tabular} \\
             \hline
(best case)  & 5                                                                  & 9.5GB & 9GB   & 9.5GB  & 3GB                                                        \\
(average)    & 2                                                                  & 3GB   & 2.8GB & 3GB    & 1GB                                                        \\
(worst case) & 0.2                                                               & 300MB & 280MB & 300MB  & 100MB                                                     
\end{tabular}
    \caption{Storage required, per node per day, assuming the node is running MD at full capacity, recording one frame per picosecond of simulation time. Values are given for different MD storage formats. The XTC format achieves greater compression because it represents atomic co-ordinates with a lower precision, so is not suitable for all use cases.}
    \label{tab:storage}
\end{table}

\begin{figure}[H]
    \centering
    \includegraphics[width=\textwidth]{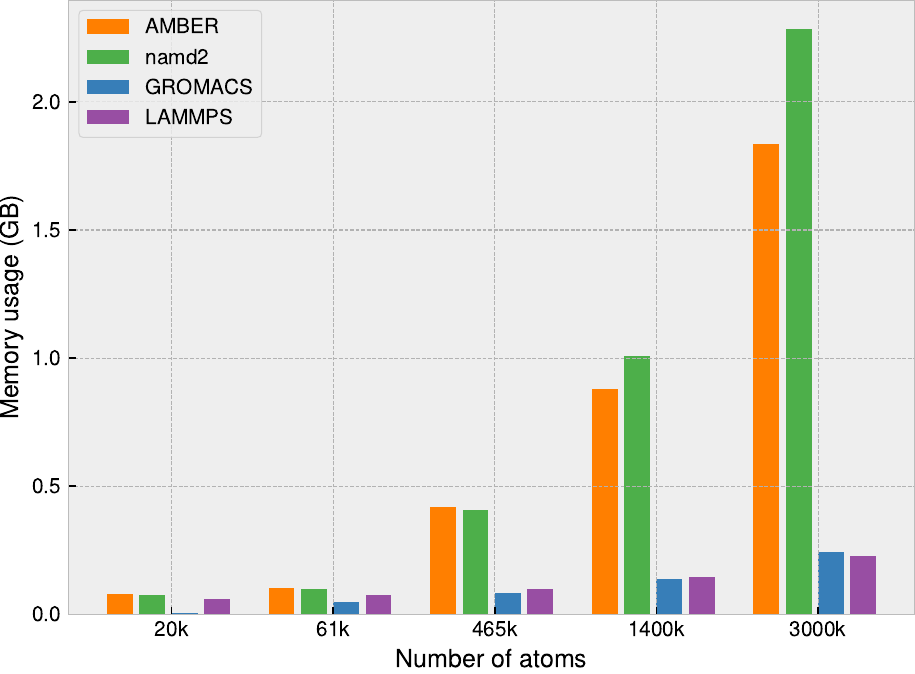}
    \caption{Molecular dynamics memory usage in GB on ARCHER2.}
    \label{fig:mdmemory}
\end{figure}

\begin{figure}[H]
    \centering
    \includegraphics[width=\textwidth]{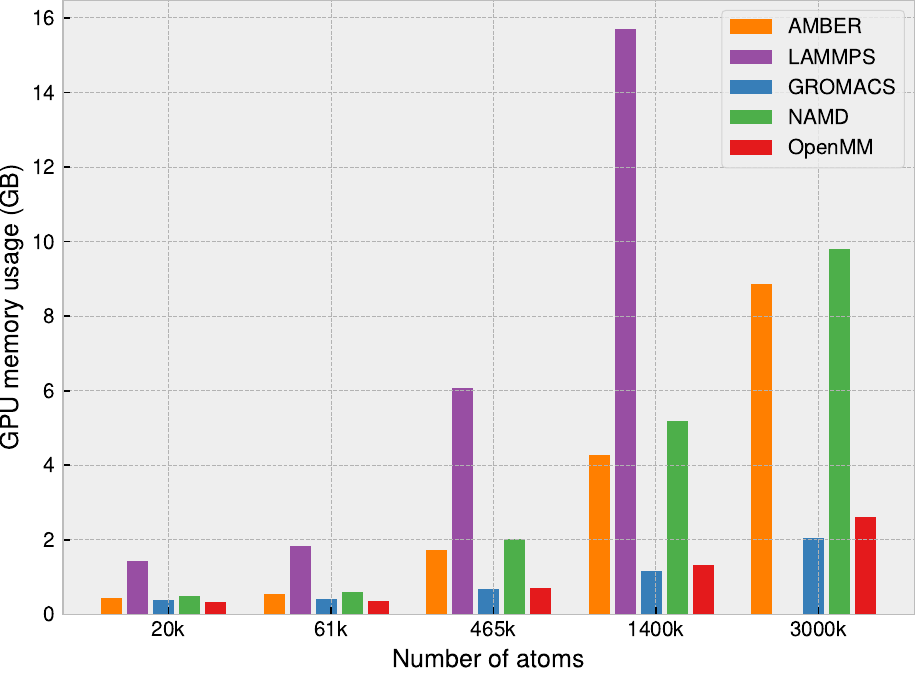}
    \caption{Molecular dynamics GPU memory usage in GB on JADE2.}
    \label{fig:mdgpumemory}
\end{figure}

Figures \ref{fig:mdmemory} and \ref{fig:mdgpumemory} show the peak memory usage of MD jobs on the CPU and GPU, running on ARCHER2 and JADE2 respectively.\\

Though there is a range of between different software, MD is rarely constrained by memory size. Data center GPUs currently start at around 40GB of VRAM, far greater than even the largest MD simulation.\\

\subsection{Multi-instance GPU}\label{sec:mig}

Nvidia GPUs from Ampere onwards can be partitioned into multiple instances, with each instance having access to some subset of the GPU's resources. These GPUs can be addressed with unique CUDA device IDs, so any software with CUDA support can run within these instances with no special configuration. For MD workloads, this can translate into an effective uplift in performance, as a single GPU partitioned into multiple simulations can reach a greater cumulative ns/day than a single simulation.\\

\begin{figure}[H]
    \centering
    \includegraphics[width=\textwidth]{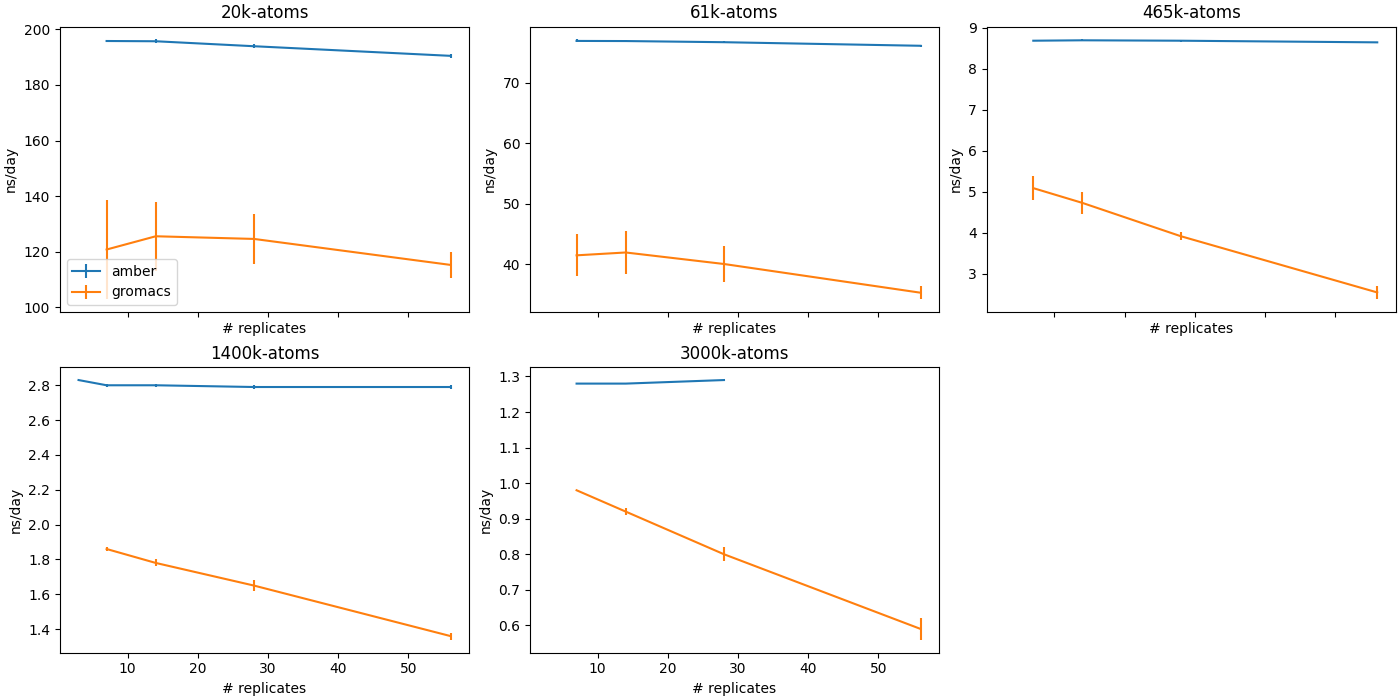}
    \caption{Multi-instance GPU benchmarks per replica for GROMACS and AMBER on an Nvidia A100, higher ns/day is better. The x-axis corresponds to the number of replicas used.}
    \label{fig:migperreplica}
\end{figure}

The MIG scaling for AMBER and GROMACS was tested on a node of Nottingham's `Ada' facility with 8 Nvidia A100 GPUs and 96 CPU cores. The node was instanced into up to 56 `virtual' GPUs. Figure \ref{fig:migperreplica} shows how the performance scales with the number of replicas used. 

AMBER easily outperforms GROMACS in terms of speed and in scaling with the number of replicas used, though it can't run more than 28 replicas without running out of GPU memory.\\

\begin{figure}[H]
    \centering
    \includegraphics[width=\textwidth]{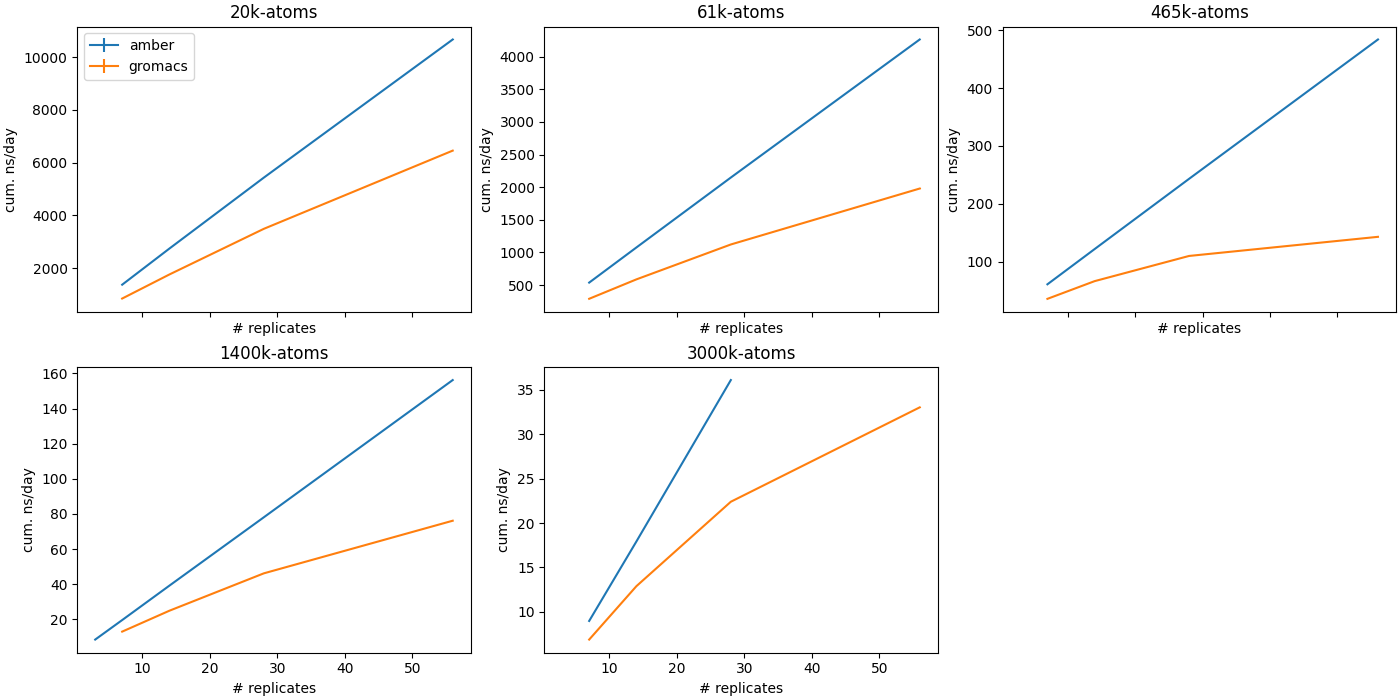}
    \caption{Multi-instance GPU benchmarks for GROMACS and AMBER on an Nvidia A100, higher total ns/day is better. The x-axis corresponds to the number of replicas used.}
    \label{fig:migcumulative}
\end{figure}

The results show that there is no performance penalty for using multiple instances with AMBER, though using GROMACS incurs a performance hit. It's likely that this reduction in performance comes not from instancing, but from the number of CPUs accessible to GROMACS\footnote{GROMACS will perform suboptimally with too few cores for a large system, but also with too many cores for a very small system, which limits the performance of the 20k atom system with few replicas.}. OpenMM and NAMD, both of which are optimised to run almost entirely on the GPU, are likely to perform similarly to AMBER.\\

As section \ref{sec:storage} shows, even the largest MD simulations can fit in just a few GBs of GPU memory, far smaller than the >40GB of workstation GPUs. Therefore, MIG provides a more efficient way for these simulations to utilise all of the resources offered by these new GPUs. MIG is transparent to users and can be enabled by HPC centres running Nvidia hardware with a few configuration flags. GPU-sharing techniques are scarcely used within the computational biology community, who tend to favour simulation speed over economy, so this presents an opportunity for training and dissemination.\\

\subsection{Coarse-Grained Molecular Dynamics}\label{sec:cgmd}

Coarse-graining is the process of merging individual atoms into groups of atoms that are represented by a smaller number of entities, thus decreasing the dimensionality and computational costs. The challenge of designing coarse-grained force fields consists of retaining emerging physical properties with a vastly reduced number of degrees of freedom in the model. It is not merely another, more efficient way of modelling biomolecular systems, but indispensable for accessing processes that occur on mesoscopic time and length scales a few orders of magnitude above the atomic scale. Coarse-grained molecular dynamics is conceptually and algorithmically very close to atomistic molecular dynamics, but has significantly reduced memory requirements due to the reduction in model complexity. Coarse-grained force fields, however, are usually bespoke interactions with intricate angular and positional dependencies to account for the loss of atomistic detail and do not comprise the simple functional forms of atomistic force fields. This entails increased computational costs, which are obviously completely offset by the gains through reducing the number of modelled entities.\\

These characteristic differences between atomistic and coarse-grained molecular dynamics have a number of practical implications. Firstly, of the MD software benchmarked in this section, only LAMMPS and more recently also OpenMM provide the structural simplicity to allow an implementation of bespoke force fields by academic teams. LAMMPS in particular provides a very mature code base, documentation and extended developer guide that enables coarse-grained force fields to be implemented in a relatively straightforward manner and benefit from functionality that would be difficult to develop within ad hoc. This strategy has been pursued for instance with the oxDNA force field, which is implemented in the LAMMPS code. Consequently, there is a tendency among coarse-grained force fields to be implemented in homegrown, user-developed codes which lack the advanced functionality and features required to obtain good performance across a variety of compute architectures. Frequent limitations include for example inferior strong or weak scaling performance on distributed memory architectures, or single-GPU support on only a subset of device manufacturers (e.g. Nvidia only). Thus, we anticipate that unless a larger and coordinated effort is made to onboard coarse-grained models into community molecular dynamics codes, many of these innovative and unique coarse-grained models will not see wider uptake. This requires a continued and improved funding for research software engineering and scientific code development.

\section{Quantum Chemistry Benchmarks}\label{sec:chemistry}

Compared to molecular dynamics, the field of quantum chemistry features a broader range of software, and also a range of methods with different computational requirements. Quantum chemistry seeks to calculate the electronic structure, physical and chemical properties of molecules. The results of these calculations are directly relevant to many biological processes \cite{merz_using_2014}, and can also be used to parameterise classical force fields for molecular dynamics \cite{wang_open_2024}.\\

\subsection{Performance}

The performance of computational chemistry calculations depends on three factors: the molecule being modelled, the set of basis functions used (which represent the atomic orbitals), and the method used to calculate the electronic structure. More complex basis sets and methods yield more accurate results, but their computational requirements have extremely harsh scaling behaviours. Two common methods are density functional theory (DFT) and coupled cluster single-double (CCSD). DFT is regarded as a computationally `cheap' method, where the CPU and memory required typically scale as $N^3$ with the system size. For CCSD, these can scale as $N^7$ with the system size.\\

The main computational expenses in these methods are matrix-matrix multiplications of extremely large matrices. Therefore, their performance is mostly limited by three factors: the CPU speed dictates the speed of the multiplications, the memory speed, latency and bandwidth dictate how quickly new values can enter the CPU cache, and the total amount of memory dictates the largest system and the most complex basis set that can be used.\\

The benchmarks in this section all use the open-source Psi4 package \cite{smith_psi4_2020}. Unlike molecular dynamics, most popular computational chemistry programs have commercial licenses or are proprietary freeware, which limits our ability to test them out and compile them for other architectures, such as ARM. Open-source options for quantum chemistry on the GPU (especially on ROCm) are particularly limited.\\

\begin{figure}[H]
    \centering
    \includegraphics[width=0.33\textwidth]{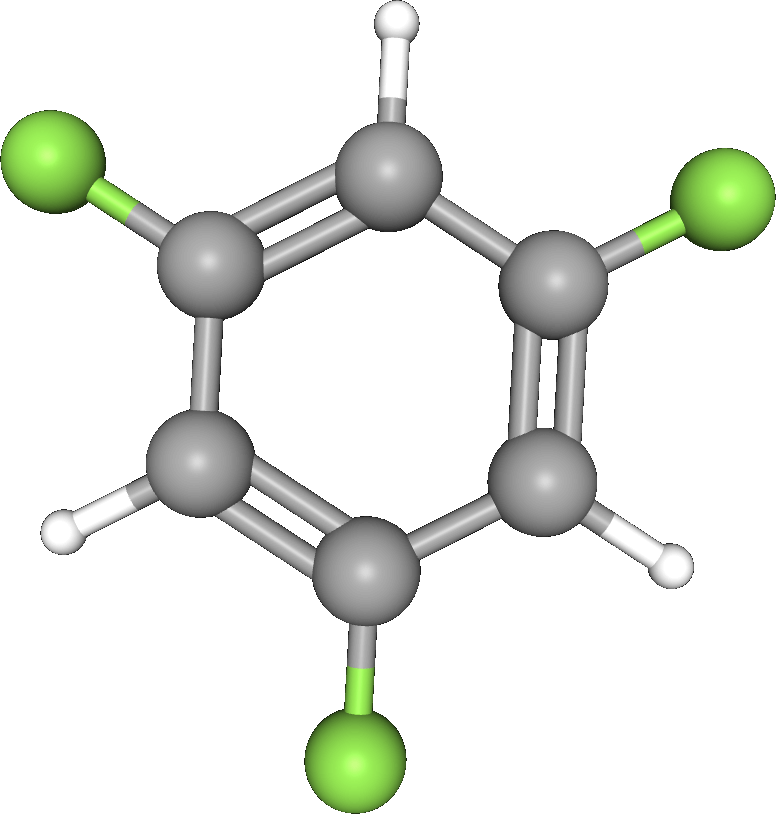}
    \caption{Trifluorobenzene molecule used as the computational chemistry benchmark system.}
    \label{fig:tfbrender}
\end{figure}

\begin{figure}[H]
    \centering
    \includegraphics[width=\textwidth]{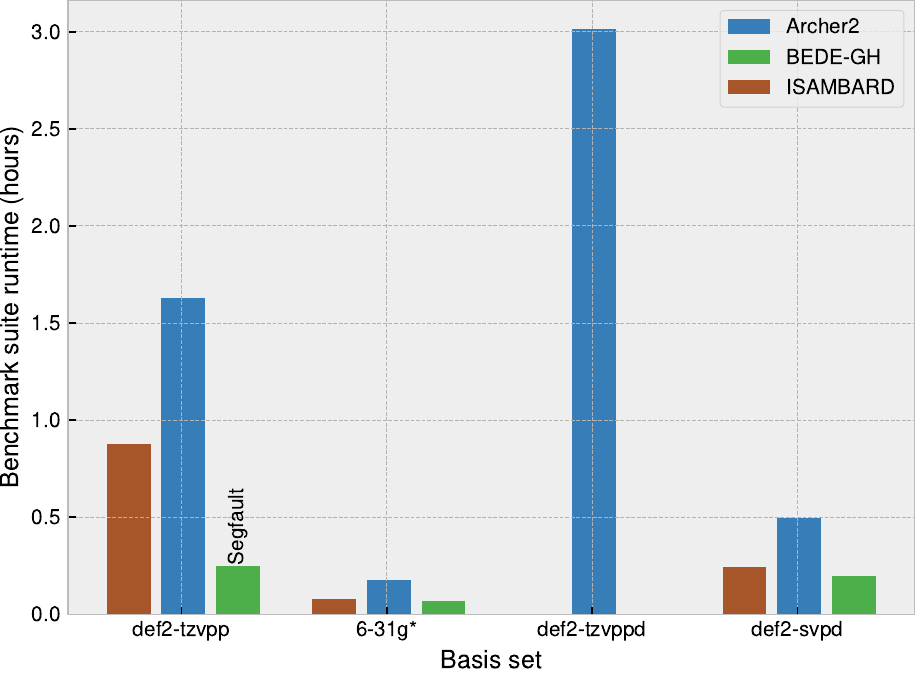}
    \caption{Psi4 benchmark suite runtime in hours, for different basis sets and on HPC different systems, using the ccsd method.}
    \label{fig:psi4time}
\end{figure}

The molecule trifluorobenzene (fig. \ref{fig:tfbrender}) is used for Psi4's benchmark suite, which calculates a range of properties using different basis sets: the Mulliken atomic charges, the L\"{o}wdin atomic charges, the electric dipole moment, the electric quadrupole moment, and the stockholder atomic multipoles. Figure \ref{fig:psi4time} shows the time taken to run the complete benchmark suite on different HPC systems.\\

JADE2 and LUMI-G were not tested, as they are GPU systems and the psi4 benchmark is CPU-only. The GH200 testbed at BEDE (and a similar one at the ISAMBARD-AI HPC facility) were tested - though these are also technically GPU systems, the extremely fast memory of the GH200 makes it an ideal candidate for quantum chemistry\footnote{These calculations are also used as part of AI forcefield training workflows, where they occur simultaneously with other work being done on the GPU}. The GH200 system performs better for simple basis sets, but with more complex basis sets, Psi4 becomes unstable. The BEDE-GH testbed system crashes while running the benchmark using the def2-tzvpp basis set, while both GH200 systems crash while evaluating the benchmark using the def2-tzvppd basis set.\\

\subsection{Memory Usage}

The memory usage of Psi4 limits the size and complexity of systems that can be used, particularly with the CCSD method. If the memory usage of a calculation exceeds the system memory, Psi4 can also utilise high-speed scratch storage, but because the performance of this scratch storage is many orders of magnitude slower than RAM, this massively increases Psi4's runtime, often beyond the time limits allocated to most HPC jobs.\\

\begin{figure}[H]
    \centering
    \includegraphics[width=\textwidth]{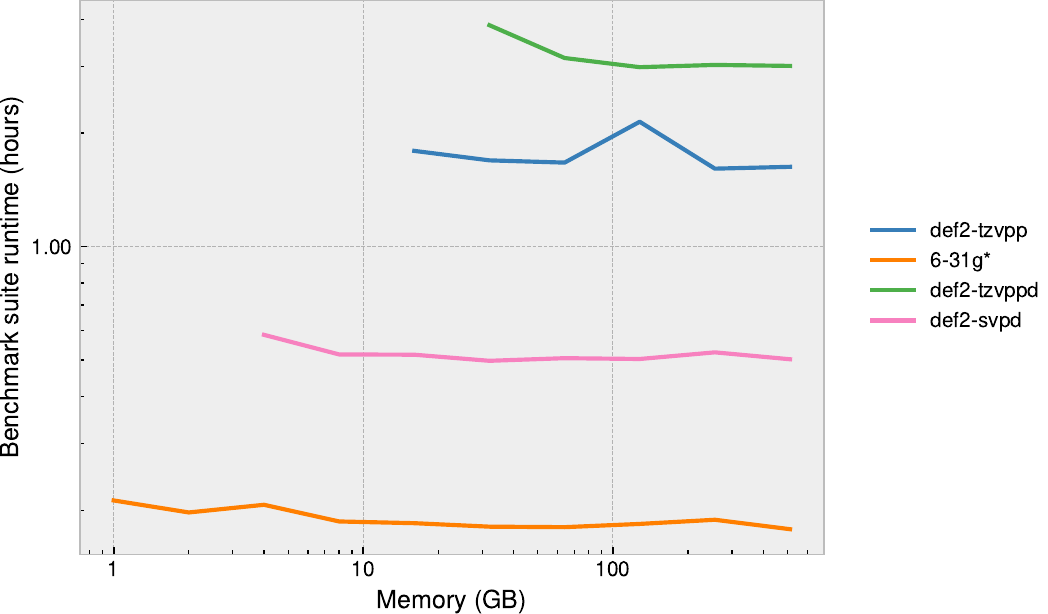}
    \caption{Psi4 benchmark suite runtime in hours, for different basis sets, using different amounts of reserved memory, on ARCHER2, using the CCSD method. Each basis set requires a different minimum allocation of memory.}
    \label{fig:psi4mem}
\end{figure}

Figure \ref{fig:psi4mem} shows the performance of Psi4 for different memory allocations. Each basis set requires a different amount of memory to run, the simplest basis set (6-31g*) runs with 1gb of RAM, while the most complex cannot run without at least 32gb of memory. The aug-cc-pV(Q+d)Z Diffuse basis set, while part of the test, could not finish within the span of a single ARCHER2 job even with the full 512gb of memory allocated. In real-world usage, particularly with larger molecules, running quantum chemistry calculations can be a balancing act, utilising simpler methods and basis sets in order to calculate properties within the limits of available memory and the speed of scratch storage. Other methods (Hartree-Fock, density functional theory, density cumulant theory, and MP4) were also tested, though these were not affected by the memory allocation. With larger systems and lower memory requirements, the performance is more determined by CPU, similar to MD.\\

\section{Electron Microscopy Data Processing Benchmarks}\label{sec:EM}

\begin{figure}[H]
    \centering
    \includegraphics[width=\textwidth]{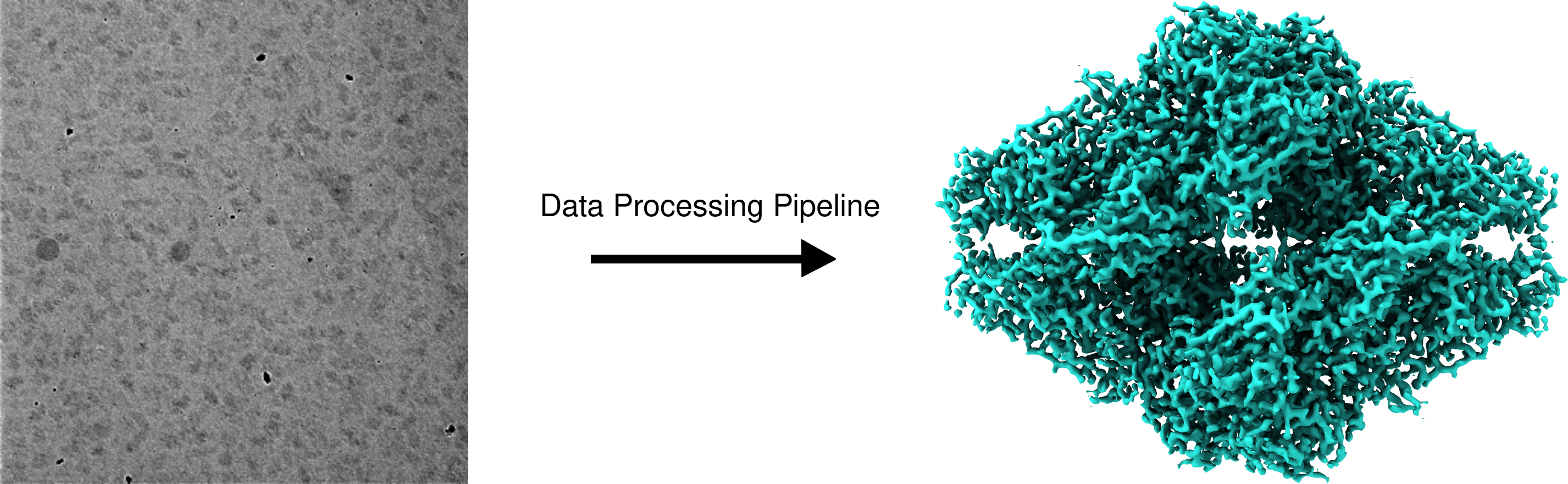}
    \caption{Before and after converting the EMPIAR-10204 dataset \cite{merk_18_2020} into an electron density map. Rendered using ChimeraX \cite{meng_ucsf_2023}.}
    \label{fig:relion}
\end{figure}

Cryogenic electron microscopy, or Cryo-EM, has allowed the structures of 120,000+  biological molecules to be determined, and is of particular use for resolving structures that are not amenable to other methods such as X-ray crystallography or Nuclear Magnetic Resonance spectroscopy.  Processing data from electron microscopes is a complex and computationally expensive task which is usually run on high-end workstations instead of HPC systems\footnote{Or on bespoke HPC systems with administrative support.}. Making Cryo-EM accessible to larger HPC infrastructure would be valuable, particularly for institutes which have limited local provisioning.\\

\subsection{Performance}\label{sec:relionperf}

Raw data from electron microscopes comes in the form of high-resolution 2D images. The process of converting these images into a 3D structure involves many steps with varying computational requirements. First, unwanted artefacts introduced by the Cryo-EM process, such as beam-induced motion, are modulated. Then, individual particles are selected from the images, and the results are further analysed to remove poorly-fitting particles and produce a set of 2D, and subsequently 3D, class averages. An initial 3D model is then reconstructed from the selected particles, and later steps in the process can refine the angular alignment of the particle images to improve the 3D model and validate the results. The initial steps (called CTF estimation and beam-induced motion correction) run only on the CPU. The particle picking steps utilise machine learning. Classification (2D and 3D) and 3D refinement steps are the most computationally-expensive, and benefit the most from GPU acceleration. In addition, the raw data is storage-intensive, usually between hundreds of gigabytes and tens of terabytes, and so requires fast temporary storage and long-term archival storage.\\

\begin{figure}[H]
    \centering
    \includegraphics[width=\textwidth]{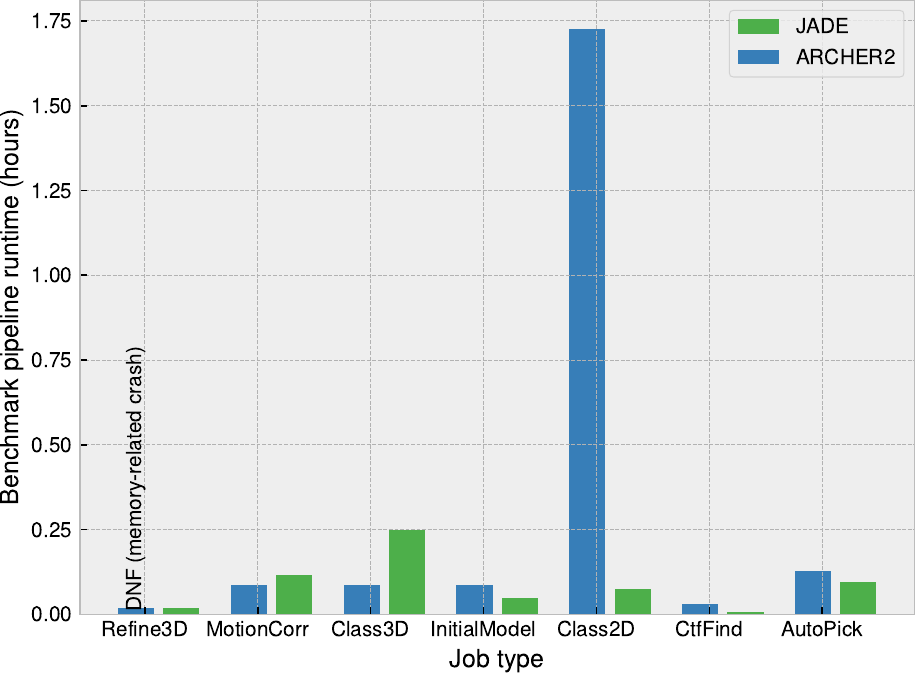}
    \caption{Relion benchmark workflow runtime in hours, for ARCHER2 and JADE2.}
    \label{fig:reliontime}
\end{figure}

Benchmarks in this section were run using RELION \cite{scheres_relion_2012, burt_image_2024} on the EMPIAR-10204 dataset, a relatively small (320GB) dataset composed of 1338 images of the enzyme beta-galactosidase \cite{merk_18_2020}. A workflow based on the RELION 3.0 tutorial was run on ARCHER2 and JADE2, the results are shown in figure \ref{fig:reliontime}.\\

The RELION SPA pipeline is a complex, multi-step process requiring several external software dependences and was found to be unstable on ARCHER2 and JADE2. On ARCHER2, Relion's performance suffers in the computationally-expensive steps of the workflow, such as 2D classification, which would normally be run via GPU \cite{kimanius_accelerated_nodate}. Meanwhile, the RELION process crashes early in the 3D refinement step due to a memory bug, rendering the benchmark unfinished. Even without these issues, the performance is further limited by a bug in the CCP-EM pipeliner tool (which manages RELION workflows), which renders the pipeliner unable to allocate more than one MPI process for some jobs.\\

On JADE2, RELION performance and stability are both greatly improved, with no crashes or pipeliner bugs. The improved performance is the result of using GPUs for the computationally-expensive steps in the pipeline. The lack of pipliner bugs is likely down to different software configuration --- JADE2 supplies OpenMPI instead of MPICH, which RELION it typically used with and has been more thoroughly tested on. JADE2's major limitation is the relative paucity of CPUs on nodes, which corresponds to reduced performance at CPU-dominated tasks, like beam-induced motion correction (MotionCorr \cite{li_electron_2013}). However, these jobs only account for a small fraction of RELION's total runtime.\\

\subsection{Compatibility with the HPC Environment}\label{sec:emcompat}

In addition to the MPI issues and crashes described in section \ref{sec:relionperf}, the HPC environment can be a harsh place, and RELION is not as hardy as most molecular dynamics software, which is used to compiling in such conditions. RELION depends upon a number of relatively common software libraries: xz, tiff, and libpng. It also depends upon a utility called ctffind4 \cite{rohou_ctffind4_2015} for CTF estimation, which in turn depends upon wxwidgets. Because HPC systems tend to provide minimal pre-installed software, none of these libraries are installed or available as environment modules, so they all have to be manually compiled. Section \ref{sec:softwareconfig} will describe steps that could be taken to make this environment more hospitable.\\

RELION also depends on several machine learning datasets, and different datasets are downloaded at configure time, build time and runtime. This is a problem on both ARCHER2 and JADE2, where FTP transfers (even incoming ones) are blocked. On ARCHER2, the easiest way to fix this is to download all of the necessary files beforehand and manually patch RELION to skip downloading them.\\

RELION was also tested on BEDE-GH, but couldn't run because ctffind4 contains inline x86 assembly. However, it is likely that EM data processing would run much faster on GH200, owing to the high memory bandwidth to the GPU, and more balanced CPU and GPU.\\

Finally, Cryo-EM is extremely storage-intense. The average EMPIAR dataset is around 2.5TB in size \cite{the_wwpdb_consortium_emdbelectron_2024}. By comparison, the default storage allocation for an ARCHER2 project is 500GB. For data transfer, ARCHER2 offers Globus/GridFTP for file transfers, while the other machines tested (BEDE, JADE2 and LUMI) only support default linux tools such as scp and rsync. Storage and data transfer are constant concerns with Cryo-EM data, but in general, storage can become a limiting factor in any HPC environment (see section \ref{sec:softwareconfig}).\\

The lack of support for RELION on HPC, alongside its various stability and compatibility issues, is indicative of a larger chicken-and-egg problem within HPC: RELION isn't commonly used on traditional HPC, RELION is not tested on HPC, RELION is hard to setup on HPC, therefore RELION is not used on HPC. Historically, HPC centres have supported a very limited selection of biophysics methods. This is largely a problem of HPC centres themselves - RELION is successfully used on modern cloud based HPC systems \cite{moriya_gotocloud_2024} and is the most widely-used EM data processing software with >122,000 depositions Electron Microscopy Data Bank (EMDB) \cite{emdb_electron_nodate}, therefore there is both opportunity and impact if additional HPC support for Cryo-EM processing can be found. Section \ref{sec:softwareconfig} suggests some steps that could be taken to broaden this support.\\

\section{Making the HPC Environment Hospitable}\label{sec:software}

So far we have only discussed the computational performance and efficiency of HPC systems. However, at some point, human beings have to use these systems --- they have to log in, administrate them, compile software, run jobs, ingest results and transfer data. The value of ns/day for a molecular dynamics simulation is only useful \textit{when a job is actually running}. This section will discuss, more qualitatively, other considerations to make sure a HPC system can be used and managed effectively.

\subsection{Nvidia and AMD}

For years, Nvidia has dominated the GPU market; most GPU programming uses their proprietary CUDA API, and 88\% of the GPU machines in the TOP500 use Nvidia GPUs \cite{noauthor_november_nodate}. Users and developers are committed to Nvidia's proprietary ecosystem, but they also prefer it to open-source alternatives, both past (e.g. OpenCL) and present (e.g. ROCm, oneAPI). CUDA is stable, well-supported and well-tested, and code written using CUDA can target a wide range of past and future Nvidia hardware.\\

AMD offers a less stable solution, with shorter windows of driver and library version support, more frequent breaking changes in updates, and less compatibility with HPC software. However, If AMD's software were to catch up with NVIDIA's, their GPUs would be a compelling alternative, as they are priced aggressively, around half as much as their Nvidia counterparts for the same level of performance.\\

The only possible compatibility issues with Nvidia come from their new Grace Hopper platform, which uses ARM CPUs. The MD software benchmarked in section \ref{sec:MD} supports ARM, but the other software described in sections \ref{sec:chemistry} and \ref{sec:EM} has compatibility issues. Psi4 compiles on ARM, though the resulting binaries are unstable, particularly in memory-intensive calculations. Most other popular quantum chemistry programs do not provide ARM binaries or source code. Relion 5 depends upon the program ctffind4, whose source code includes inline x86 assembly, and so cannot be compiled on ARM. Finally, many popular scientific python packages are missing or broken on ARM, forcing users to build them from scratch\footnote{One dependency of Relion 5 --- libigl --- supplies an empty package instead of a source distribution, which results in a broken build without manual fixes.}. These kinds of problems are common on ARM, and porting scientific software to Nvidia's new accelerator cards will require additional work from system administrators and software authors.\\

In comparison, AMD's accelerator cards use x86\_64 CPUs, averting these issues entirely. However, using ROCm brings its own challenges. Support for AMD is incomplete, even in established scientific software. Only four of the five benchmarked molecular dynamics programs support ROCm at time of writing (previous versions of OpenMM were supported, but OpenMM-HIP's developers are still working on supporting the current version). RELION 5 has ROCm build options, though its main competitor, CryoSPARC, does not. Quantum chemistry software support for GPUs is less common, but ROCm support in particular is very limited: CP2K has partial support, and some other software (CASINO and Octopus) will run on AMD hardware via older software libraries (OpenCL and OpenACC).\\

For developers, adding ROCm support is easy but not automatic. AMD provides semi-automated tools to convert CUDA code to ROCm, but the process still requires manual tuning (plus the continuous work needed in the long-term to support new ROCm versions). The largest cost incurred by developers, system administrators and users on ROCm/HIP comes in the form of ROCm software updates. Because ROCm is a relatively new API, features are frequently added and changed, and developers often have to update their applications to deal with deprecated functionality and API changes. Moreover, each AMD driver only supports a limited set of ROCm versions ($\pm$2 minor versions each way). This means that updating software often requires a ROCm driver update, and vice versa. Inevitably, this narrow window means that some software will fall through the cracks, stuck outside the range of support provided by drivers and system software. ROCm user-level libraries contain hardcoded references to the default install paths, meaning users can't easily install their own versions of the ROCm software. New versions of ROCm can also come with unexpected problems --- with the release of ROCm 6, most LUMI-G users saw enhanced performance, but some also reported significant performance regressions. These regressions were later fixed in ROCm 6.2.\\

Numerical stability is also worth considering. Most MD ROCm implementations are very new, and aren't as well-tested as their CUDA counterparts, so they can be more prone to bugs and numerical errors. Table \ref{tab:conservation} shows whether the total energy of the system was conserved during benchmark runs. If energy isn't conserved, the numerical integrator is unstable for that combination of hardware and software.\\

\begin{table}[H]
\centering
\begin{tabular}{lllll}
            & AMBER     & GROMACS     & NAMD      & LAMMPS    \\
            \hline
20k atoms   & \checkmark & \checkmark   & \checkmark & \checkmark \\
61k atoms   & \checkmark & \checkmark   & \checkmark & \checkmark \\
465k atoms  & \checkmark & \checkmark   & \checkmark & \checkmark \\
1400k atoms & \checkmark & \textbf{X} & \checkmark & \checkmark \\
3000k atoms & \checkmark & \checkmark   & \checkmark & \checkmark
\end{tabular}
    \caption{Conservation of energy running molecular dynamics on the MI250x on LUMI.}
    \label{tab:conservation}
\end{table}

The energy of the 1400k atom simulation on GROMACS is subject to random spikes, which indicates numerical instability. This result doesn't fully exonerate the other simulations --- the conservation of energy is a coarse measure of simulation accuracy and does not guarantee that the other simulations listed in table \ref{tab:conservation} are free from numerical problems.\\

All of these issues with AMD's software are ultimately solvable, and the situation should improve as it becomes more mature and more adopted. Furthermore, there are some genuine advantages to AMD's methods --- in particular, ROCm/HIP are distributed under an open source license that should help to ensure a greater degree of interoperability between GPU-enabled software and hardware, and decrease the risk of vendor lock-in. However, in short term, ensuring that this hardware is consistent and usable will require greater user, developer and system support. Procurement decisions regarding AMD GPUs and ARM CPUs in particular should be made holistically, taking into account the investment in staffing required to support that hardware.\\

\subsection{Software Configuration}\label{sec:softwareconfig}

\textbf{Several of the tested HPC systems suffer from unusually long periods of downtime} --- for example, during 2023, ARCHER2 underwent a major software update, and was unavailable between the 19th of May and the 12th of June. Combined with several other periods of downtime, ARCHER2 was offline for 28 days that year, with an uptime of 92\%. In other fields of Linux server administration (e.g. cloud computing, data storage, web servers) it's common for service providers to agree to provide a certain amount of uptime, commonly measured in 'nines of uptime'. Three nines, or 99.9\% uptime, is considered standard. In comparison, in 2023, ARCHER2 achieved one nine. In particular, HPE Cray Machines are prone to these protracted outages, with LUMI down for 28 days continuously due to a software upgrade in 2024 (other outages during 2024 mean that LUMI did not achieve a single nine in that year). The uptime of these systems could be improved with phased maintenance, which is the norm in big datacentres. This approach is more labour-intensive and would require a greater investment in staff, but this is far outweighed by the increased uptime and lack of disruption to users.\\

\textbf{HPC systems often lack necessary software and system libraries, and also lack the tools needed to build that software from source} --- HPC systems typically provide minimal Linux installations, with a limited selection of scientific software, system libraries, compilers, a selection of maths libraries, and build tools. Often, these tools and libraries are outdated (for example, JADE2 stopped providing new versions of most MD software after 2020). Often, the provided versions of these tools aren't fit for their intended purpose --- for example, ARCHER2 provides a LAMMPS module, but it's a minimal build of LAMMPS, with most of the optional packages not included. Many users instead opt to compile their own software, which can be time-consuming, and they often need to contend with outdated or missing system libraries and outdated compilers.\\

There's no quick fix to these problems --- adding more software, or updating the software more regularly would also increase the maintenance burden of the system. However, there are other strategies to help mitigate these problems and save users and system administrators time:
\begin{itemize}
\item Offer software via EasyBuild or Spack --- this reduces the maintenance burden on system administrators, and makes it easy for users to compile software from source. For example, LUMI-G provides recipes for hundreds of pieces of common scientific software via EasyBuild.
\item Offer some form of virtualisation or containerisation --- this allows users to pull software from existing container image libraries, and configure their systems in a more granular way, with more control over libraries and configuration. For example, ISAMBARD-AI provides podman-hpc, which allows containerised applications to achieve a high level of performance and integrate with common HPC libraries and tools like MPI and CUDA.
\item Begin exploratory work into flexible provisioning of HPC resources using tools such as OpenStack --- this solution would allow for HPC resources to be provisioned like cloud computing resources, similar to Azure or AWS, which might suit the needs of some users better than a traditional monolithic system with a SLURM queue.
\end{itemize}

\textbf{Work is duplicated between HPC centres} --- Many of the issues described in previous sections stem from a lack of time and resources within HPC centres. While technical solutions (like build frameworks and containerisation) can help, this is also a human problem with possible human solutions. Time and energy could be saved by sharing knowledge, expertise and methods between HPC centres. For example, most HPC systems in the UK implement bespoke account management, login systems and 2-factor authentication (2FA), and many run on  their own instances of EPCC's SAFE account management software. Collaboration between HPC centres across the board could help them to develop common re-usable processes for installing and deploying software and managing updates, and improve both uptime and software availability across the board.\\


\textbf{Storing and transferring data to/from HPC systems is often difficult and time-consuming} --- As discussed in sections \ref{sec:storage} and \ref{sec:emcompat}, the amount of data that needs to be stored on and transferred to/from these machines will only ever increase. Storage needs to be provided for both short-term and archival use --- however, currently, it isn't clear whether this is the responsibility of HPC centres or universities or individual research groups. The answer to this question is beyond the remit of this report, but it does need to be answered before the next generation of HPC systems are built. Transferring data is an equally thorny issue --- of the machines tested for this report, only one (ARCHER2) had a Globus endpoint, none supported Aspera, and most blocked connections on port 20, disallowing downloads via FTP.\\

\textbf{Efficient HPC usage is viewed as being less important than raw speed} --- the majority of the existing benchmarks for molecular dynamics are not efficiency-minded and instead focus on juicing raw ns/day values or system size as high as they can go. Scaling to hundreds of nodes, multiple GPUs or both are more represented than single-node/single-GPU performance. Even HPC queues are set up with this type of scaling emphasised --- ARCHER2's standard QoS allows users to submit jobs that run on up to 1024 nodes at once, but the maximum job length is 24 hours, even though running a single molecular dynamics job on fewer nodes for longer would produce a many-fold improvement in efficiency. Furthermore, this QoS also limits the maximum of jobs that can sit in the queue at once to 64. These queue settings encourage users to scale up fewer jobs to many nodes, instead of running longer jobs, or running more jobs\footnote{Other QoS options are available on ARCHER2, though they only marginally extend the limits on job quantity and length, and they can't be used together (ARCHER2 users have the option of doubling the number of jobs they can have in the queue, or submitting jobs of up to 96 hours, but not both).}.\\

\textbf{Sustainable HPC is a people problem as much as it is a hardware/software problem} --- this report outlines many dramatic changes that will occur in the landscape of HPC over the next few years --- new hardware architectures, new software, new use cases, and new ways for users to interact with their machines. All of this will incur an additional cost in training and expertise, from users to developers to HPC staff. Our ability to adapt to these changes and build machines that actually work depends on our ability to hire and retain high-quality system administrators.



\section{Conclusion}\label{sec:conclusion}

The biomolecular simulation landscape has evolved significantly in the last decade, with the emergence of a diverse array of numerical and physical modelling methods. This evolution has been driven by advancements in compute capability alongside innovation in experimental methods, which has led to a step change in the data available for simulation studies at the same time as a step change in compute, unlocking enormous potential for new science. This increase in compute capability brought with it increased diversity in hardware platforms and their supporting software stacks, and so has introduced a layer of complexity into the landscape of scientific software and tooling. Table \ref{tab:summary} summarises the different computational requirements for commonly used methods in the domain.\\

\begin{table}[H]
\centering
\begin{tabular}{llllllll}
                    & Memory                          & CPU    & GPU       & Storage    & ARM & AMDGPU          \\
                    \hline
                    
MD  & \textless{}=10GB          & High   & Yes  & $\sim$100GB/dataset       & Yes          & Yes* \\
QC   & 10-512GB & Medium & None       & $\sim$10MB/dataset                                 & No           & No                         \\
EM & \textgreater{}10GB     & High   & Yes       & $\sim$2.5TB/dataset & No           & Yes* \\
CGMD  & \textless{}\textless{} 10GB  & High & Yes  & $\sim$100GB/dataset  & Yes & Yes                       
\end{tabular}
\caption{Summary of the computational requirements of different methods.}
\label{tab:summary}
\end{table}

Handling this complexity is an ongoing concern for all parties. At HECBioSim\footnote{HECBioSim is the HPC-focussed consortium responsible for the efficient distribution of compute at UK Tier1 and UK Tier2 HPC facilities, to the research community in biomolecular simulation}, we have seen this play out over the last decade of allocating resources --- the increased diversity of hardware on the machines available and the diverse array of simulation methods employed by the wider community leads to confusion amongst the user-base about which methods work best and on which hardware. HECBioSim currently mitigates this by technical support and allocation of resources based on the technical suitability of codes for various hardware platforms, which is informed by benchmarks. This approach has taken overall utilisation from around 70\% up to around 97\% of the consortium allocation on HPC facilities. The other major functionality that HECBioSim serves is to provide front-line specialist software support and training in the best-practice use of HPC for biomolecular simulation. This has been used to spectacular effect on JADE2 and Bede, where software packages for biomolecular simulation were compiled and maintained by the HECBioSim CoSeC support. This cooperation between the computing centres and the expert community reduced the workload on stretched core system administration teams while providing domain specific software optimisation and frontline support. During the COVID-19 pandemic, the true value of this approach was seen, when large amounts of compute could be directed toward front-line research without huge amounts of technical debt as a barrier to exploitation. There are only 6 HECs and not all have funded CoSeC experts supporting HPC, HECBioSim is notably ahead in this area.\\

\textbf{OPPORTUNITY: continue funding the consortium model for HPC communities, broadening this model to other communities such that they manage their own software.  Training users in best practice and the nuances in how software and hardware interact will take significant pressure away from core system administration teams at computing centres.}\\
 
The next generation machines (e.g. MI300, GH200) are likely to pose even greater complexity. The current generation of hardware involves several CPU architectures based on x86, Arm or POWER with the GPU architectures largely dominated by various generations of Nvidia hardware. The next generation hardware already has a similar mix of CPU architectures available from vendors but the GPU options available now see additions from AMD and Intel, along with novel chips from Nvidia, with a number of platforms that combine CPUs and GPUs in different permutations. Though these new platforms are often billed as machine learning devices, in reality they are general purpose computers that offer excellent performance in a wide range of applications, including molecular dynamics. Machine learning is just one part of the larger scientific computing ecosystem, and is normally used in conjunction with (and is often trained from) physics-based models, this notion of special hardware for special purposes is largely an artificial distinction drawn at the software stack level. There is an opportunity here to maximise investment in such hardware by broadening participation. However, supporting this new hardware will require more work from system administrators, user support, and users themselves.\\

\textbf{OPPORTUNITY: broaden participation in established next generation machines to support the underpinning physical models behind the development of AI technologies.}\\

AMD GPUs are a compelling alternative to Nvidia, due to their lower price and open software ecosystem. AMD’s ROCm is now supported by most scientific software, however, as with AI accelerators, using AMD GPUs will require additional investment in system administration and user support, particularly while the AMD software ecosystem is still young. Table \ref{tab:features} shows an overview of the software and hardware compatibility of different methods on old and new hardware and software.\\

\begin{table}[H]
\centering
\begin{tabular}{llllllll}
        & \multicolumn{5}{l}{Molecular Dynamics}             & QMM         & EM          \\
        & Gromacs & Amber      & NAMD & LAMMPS  & OpenMM     & Psi4        & Relion      \\
         \hline
x86 & \checkmark       & \checkmark          & \checkmark    & \checkmark       & \checkmark          & \checkmark           & \checkmark           \\
ARM     & \checkmark       & \checkmark          & \checkmark    & \checkmark       & \checkmark          & \checkmark *          & \textbf{X}  \\
MPI     & \checkmark       & \checkmark          & \checkmark    & \checkmark       & \textbf{X} & \textbf{X}  & \checkmark           \\
OpenMP  & \checkmark       & \textbf{X} & \checkmark    & \checkmark       & X*          & \checkmark *    & \checkmark           \\
CUDA    & \checkmark       & \checkmark          & \checkmark    & Partial & \checkmark          & \textbf{X} & \checkmark           \\
ROCm    & \checkmark       & \checkmark          & \checkmark    & Partial & X*         & \textbf{X}  & \checkmark           \\
        &         &            &      &         &            &             &            
\end{tabular}
    \caption{Feature support from different software packages. Psi4 supports ARM, but with a lower level of overall stability, and OpenMP but only for some tasks. OpenMM only supports ROCm through a third-party addon with a slower release schedule. LAMMPS support for both CUDA and ROCm is partial. OpenMM supports OpenMP only for its reference implementation, which is not recommended for general use.}
    \label{tab:features}
\end{table}

The work done in this study and the resulting information generated presents a comprehensive look across the software landscape in the field, but there are still some nuances that were beyond the scope of this study, due to the time limitations and the breadth of work already planned. This study only looked at the main algorithms that are commonly used for simulation purposes. These software packages are often over a million lines of code and there are situations where features outside of these benchmarks present a worse picture of software readiness than detailed here. For example we know that in the AMBER MD engine, not all features are available in the high performance PMEMD implementation for HPC, and certainly not all features available in the MPI compilations are available in the CUDA compilations. This is true across all of these tools -- vital feature sets in these codes are not all available across the whole of the hardware landscape; this represents more work to be done at the software engineering level to bring these scientific features to modern HPC.\\

\textbf{OPPORTUNITY: further investment in scientific software programmes that focus on the maintenance and advancements of existing codes. Continued investment in the CCPs would be welcome. Recent expansion into areas of UKRI by the DRI not covered by the EPSRC investment in CCPs is a healthy sign.}\\

Computing centres running HPC facilities face the realities of running the hardware and maintaining the core system software stack as well as having to support the user communities. This is a difficult undertaking with the current funding landscape. According to proceedings at the CIUK supercomputing conference in December 2024, teams working at this interface are reporting significant pressures recruiting and retaining talent. The reasons reported were diverse: the specialist multidisciplinary niche, meaning there is no single path to prepare for a career in HPC systems administration and training new system admins is a lengthy process. The constant loss of staff to the private sector, such as organisations that specialise in hosting and/or using cloud computing services or moving into adjacent careers such as software development. Running HPC facilities is a constant balancing act between the developing software stacks of the core system libraries and the developing software across many research domains. Maintaining compatibility across this landscape often involves holding software versions stable for long timescales, which means system administration is often lobbied by various communities, some of which require the bleeding edge, and others requiring older but more stable features. To compound this, communities range in cultural approach to software: some communities, like biomolecular simulation, have a handful of million line codes that users expect to be present and optimised on systems for them to begin doing science, whilst other user-groups are made up of many single developer projects where often the user is the developer, and they want access to much more control over software. To support all of these requires a different approach to each. There are technical solutions to all of this, such as software build tools like EasyBuild/Spack or virtualisation technologies involving containerisation like Singularity/podman-hpc. There is an opportunity here --- increased cooperation around software tools and build systems would take pressure away from core systems teams. The other option is greater levels of funding for HPC facilities to recruit new talent with these skills or to train existing teams in these newer technologies.\\

\textbf{OPPORTUNITY: Increase cooperation around software build systems and package distribution. Greater cooperation between the user communities, RSEs at institutions, CoSeC specialists and computing centres would shift the pressure from being focussed on core systems teams alone.}\\

Storage and data transfer remains a difficult issue. Supercomputers should have storage provisioned according to their computing power. A new Tier 1 machine could produce tens or hundreds of petabytes of data per year, and the computational power that such a machine offers is useless without a way to store the data it produces. So far, only the PSDI BioSimDB project has attempted to tackle this problem, though this only provides 2PB of archival storage. Storage capabilities in the UK physical sciences lag significantly behind the capabilities in compute, both in the capacity, speed of transfer inter-institution and in long term centralised storage. \\

\textbf{OPPORTUNITY: Ensure other investments in data infrastructure by the EPSRC and UKRI are embedded into the communities focussed around HPC.}\\

This study has looked at the readiness of a wide-reaching part of the software landscape across a representative sample of architectures that are likely to form parts of any future HPC distribution. Our findings then are broadly in agreement with the \href{https://www.gov.uk/government/publications/future-of-compute-review}{future of compute review} \cite{noauthor_independent_2023}. We have identified that there are many existing programmes where the opportunity is to simply continue them, if not to expand them across to other domains within council remit areas. We have identified that there are opportunities to solve some of the issues faced by the collective HPC communities, by organising and cooperating research groups, HPC centres, RSE teams and programmes like CoSeC. Investment in this area could be as simple as providing the means to co-locate teams for short periods, run workshops/hackathons/retreats and establish an event at the CIUK to showcase outcomes and network.\\

The purpose of this study has not been to write an extended business or science case for exascale computing, because this has been done elsewhere \cite{mulholland_large-scale_2024, noauthor_independent_2023, wilkinson_ukri_2021}. Instead it is to assess the next generation platforms and their associated complexity against the readiness of our scientific codes, to establish routes to exascale. This document has demonstrated that our community is one of the leading communities in this regard. The workloads that our community represents are not simple to classify into any particular flavour of hardware architecture, we have methods such as coarse-grained molecular dynamics that scale better on CPUs whilst all-atom molecular dynamics scale well on GPUs. AI brings with it new challenges, especially at the interface with physics based methods like ours. There is no simple answer to the question `which platform is best?', there are caveats to each and every approach, with work involved for different parts of the community depending on what is chosen. An exascale landscape involving a diverse range of hardware and teams would be more favourable than a single architecture machine over an exaflop in size with a single team to manage it. Through problems identified in this study, we have shown that if we empower the whole community through investment in people to work together around investment in hardware, we can bring step-change in how UK HPC is delivered. \\

\newpage

\section{Contributors}
Report authors:
\begin{itemize}
\item Figures/benchmarking/method/software development (except section \ref{sec:mig}) by Rob Welch
\item MIG Benchmarks and figures (section \ref{sec:mig}) by Charles Laughton
\item Report text:
\begin{itemize}
    \item Sections \ref{sec:intro}-\ref{sec:software}, except section \ref{sec:cgmd} by Rob Welch
    \item Section \ref{sec:conclusion} by James Gebbie-Rayet, Rob Welch
    \item Section \ref{sec:cgmd} by Oliver Henrich
\end{itemize}
\item Editing: James Gebbie-Rayet, Tom Burnley, Daniel Cole, Alan Real, Sarah Harris, Mark Wilkinson
\end{itemize}
HPC Facilities used:
\begin{itemize}
\item We acknowledge the EuroHPC Joint Undertaking for awarding this project access to the EuroHPC supercomputer LUMI, hosted by CSC (Finland) and the LUMI consortium through a EuroHPC Regular Access call. (thanks Kurt from LUMI-G support!)
\item This work used the ARCHER2 UK National Supercomputing Service \cite{beckett_archer2_2024} (thanks Arno at ARCHER2 support!)
\item This work made use of the facilities of the N8 Centre of Excellence in Computationally Intensive Research (N8 CIR) provided and funded by the N8 research partnership and EPSRC (Grant No. EP/T022167/1). The Centre is co-ordinated by the Universities of Durham, Manchester and York (thanks to Mark Dixon and the BEDE team at Durham).
\item ISAMBARD-AI (thanks to Chris Woods, Lester Hedges, and the ISAMBARD team at Bristol)
\item Ada at Nottingham
\item JADE2
\end{itemize}
Input files:
\begin{itemize}
\item HECBioSim benchmark input files provided by James Gebbie-Rayet
\item Relion input files provided by Matthew Iadanza (help via Tom Burnley, Joel Greer)
\item Psi4 input files provided by Charlie Adams (help via Edina Rosta, Daniel Cole and Phil Hasnip)
\end{itemize}
ExaBioSim team: Syma Khalid, Agnes Noy, Julien Michel, Antonia Mey, Shozeb Haider, Davide Marenduzzo, Jonathan Essex, Rosana Collepardo-Guevara, Franca Fraternali, Dmitry Nerukh\\
EPSRC funding via the ExCALIBUR programme EP/Y008693/2

\newpage

\bibliography{lib} 
\bibliographystyle{unsrturl}

\section{Corrigendum}\label{sec:corrigendum}

Some of the simulations listed in section \ref{sec:MD} used values for simulation parameters that were not consistent between different molecular dynamics software. These are:

\begin{itemize}
\item The OpenMM simulations were run without pressure coupling.
\item The cutoff distance is not specified in the AMBER input files and defaults to 8{\AA}, instead of the 12{\AA} used in the other benchmarks.
\end{itemize}

These errors have been corrected for the input files hosted on the HECBioSim website. While they don't affect the conclusion of section \ref{sec:conclusion}, they do mean that AMBER's performance and efficiency are overstated in the plots of section \ref{sec:MD}.\\

In addition to these errors, section \ref{sec:methodology} describes several cases in which direct, like-for-like comparisons between MD software are not possible.

\section{Appendix}

\subsection{Benchmark Methodology}\label{sec:methodology}

The benchmarks described in this report were run using a utility named hpcbench, which is available on GitHub at \url{github.com/HECBioSim/hpcbench}. This utility provides tools to generate SLURM submission scripts from templates, monitor and log information from running HPC jobs, package that information into .json text files, and plot results from those text files. The job submission scripts can be found within the hpcbench repository, in the \texttt{hpcbench/job\_scripts} directory. The input files for each benchmark can be found at the HECBioSim website, \url{hecbiosim.ac.uk/access-hpc/hpc-benchmarking-suite}, and the raw data for all the results can be found in the repository \url{github.com/HECBioSim/benchmark-results}. Note: the raw data is in the hpcbench json log format. The \texttt{sample\_plots.sh} shell script inside the benchmark results repository contains the commands necessary to recreate the plots from this report using hpcbench. The README.MD file for hpcbench provides more general info on how to use the hpcbench's plotting and analysis tools.\\

Benchmarks were configured and run in a way that is representative of the user experience of an average user on a given machine, and \textit{not} to maximise performance or to rigorously control for differences in software configuration on each machine. This usually means compiling and running software in line with the official documentation for each HPC center and piece of software. For example, the GROMACS input scripts on ARCHER2 are derived from the official ARCHER2 documentation for GROMACS (\url{docs.archer2.ac.uk/research-software/gromacs}). If environment modules for the software were provided in the HPC environment, those were used instead of compiling from scratch. The versions of software and libraries used on different machines are shown in table \ref{tab:versions}.\\

\begin{table}[H]
\centering
\begin{tabular}{lllll}
         & JADE2            & ARCHER2  & BEDE-GH  & LUMI-G         \\
         \hline
GROMACS  & 2022.2           & 2022.4   & 2024     & 2024.1         \\
AMBER    & amber20          & amber22  & amber22  & amber24        \\
NAMD     & 3.0-alpha7       & 2.14     & 3.06b6   & 3.0            \\
OpenMM   & 8.1.1            & n/a      & 8.1.1    & n/a            \\
LAMMPS   & 2Aug2023         & 2Aug2023 & 2Aug2023 & 2Aug2023       \\
Psi4     & n/a              & 1.9.1    & 1.9.1    & n/a            \\
RELION   & 5                & 5        & n/a      & n/a            \\
Compiler & gcc 9.1.0        & gcc 11.2 & gcc 13.2 & AMD clang 17.0 \\
CUDA     & 9 (OMM:11.4) & n/a      & 12.3.2   & n/a            \\
ROCm     & n/a              & n/a      & n/a      & 6.0.3         
\end{tabular}
\caption{Software versions used for benchmarking}\label{tab:versions}
\end{table}

The HECBioSim benchmark suite was used for benchmarking all molecular dynamics software. As with other aspects of the benchmarking methodology, this suite was designed to be representative of of the kinds of simulations that computational biologists actually run. GROMACS, AMBER, NAMD, LAMMPS and OpenMM implement very similar features, but they don't have exact feature parity with one another. For example, GROMACS resolves constraints using the SETTLE algorithm, while LAMMPS doesn't implement SETTLE, and the LAMMPS benchmark uses the more computationally expensive SHAKE algorithm instead. The OpenMM benchmarks use the LFMiddle discretisation for their dynamics, which is the only discretisation supported by newer versions of OpenMM. The HECBioSim benchmarks are as close to parity as possible, but it's not possible to make these programs to recreate exactly the same physics. Full input files are available in the benchmarks repository, but the most important parameters are: all MD simulations were run using the CHARM22 force field (except OpenMM, which used CHARMM36), at 300K, with a timestep of 2 femtoseconds. The three larger systems ran for 10000 steps, while the two smaller ones ran for 50000 steps.\\

The quantum chemistry benchmarks utilise Psi4 and the CCSD method exclusively, with the trifluorobenzene molecule, which is comparatively small. This does not completely reflect the full range of quantum chemistry software or methods. This decision stemmed from practical limitations; a lot of quantum chemistry software is commercial or proprietary freeware, which limits its availability and portability for this project. Other quantum chemistry methods, such as density functional theory, have similar computational requirements to atomistic molecular dynamics, so CCSD was chosen specifically because of its higher memory requirements, to ensure that the benchmarks would represent users running memory-intensive jobs.\\

For the EM data pipeline, the RELION 3.0 benchmark is used. This is a much smaller dataset than most EM jobs, but the RELION benchmarks were constrained by the limited storage on most HPC systems alongside the need for a fully tested and reproducible RELION pipeline. The RELION benchmark input files were modified based on the system they were being run on, and the different versions of each file can be found in the HECBioSim benchmarks git repository (\url{github.com/HECBioSim/benchmarks}) with the job script template files in \url{github.com/HECBioSim/hpcbench}. 

\subsection{Energy Conservation on LUMI-G}

\begin{figure}[H]
    \centering
    \begin{minipage}{0.45\textwidth}
    \centering
    \includegraphics[width=\textwidth]{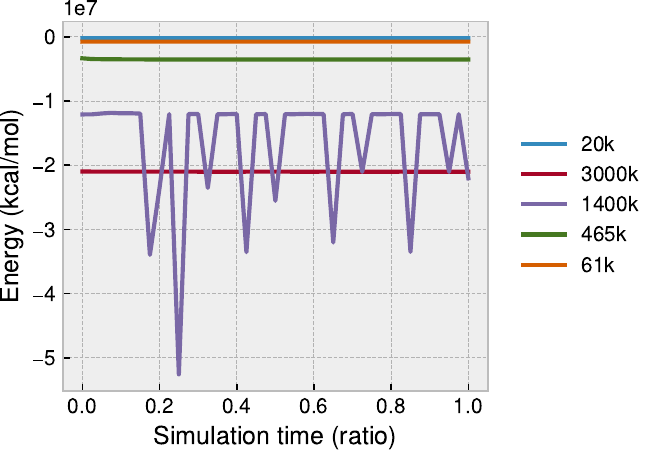}
    \caption{Conservation of energy in GROMACS running on LUMI-G.}
    \label{fig:gromacsenergy}
    \end{minipage}\hfill
    \begin{minipage}{0.45\textwidth}
    \centering
    \includegraphics[width=\textwidth]{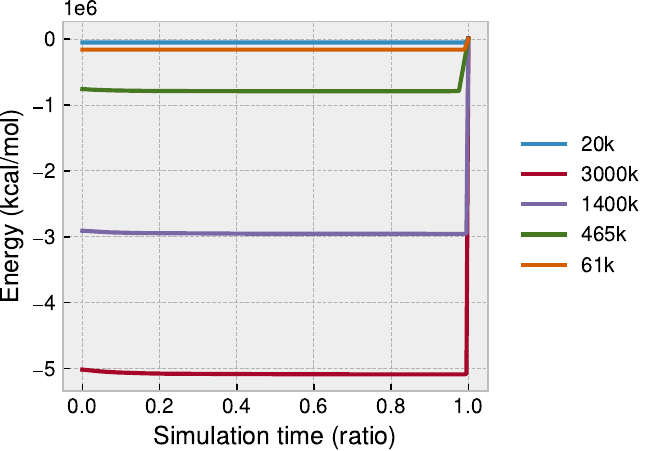}
    \caption{Conservation of energy in AMBER running on LUMI-G.}
    \label{fig:amberenergy}
    \end{minipage}
\end{figure}

\begin{figure}[H]
    \centering
    \begin{minipage}{0.45\textwidth}
    \centering
    \includegraphics[width=\textwidth]{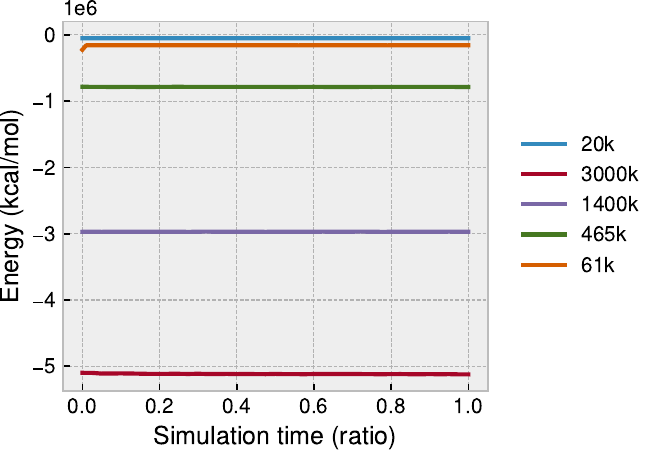}
    \caption{Conservation of energy in NAMD running on LUMI-G.}
    \label{fig:namdenergy}
    \end{minipage}\hfill
    \begin{minipage}{0.45\textwidth}
    \centering
    \includegraphics[width=\textwidth]{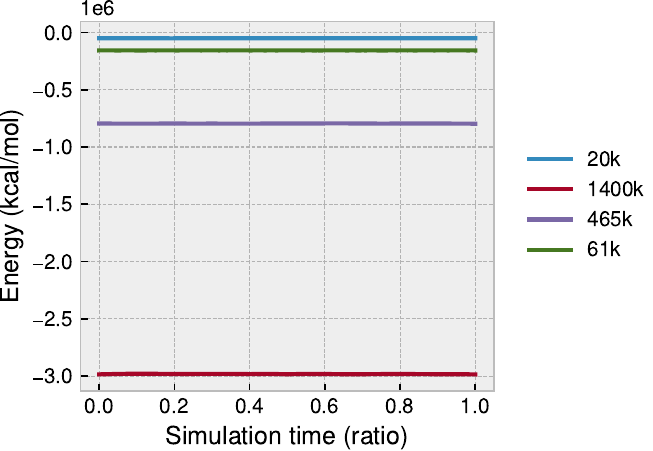}
    \caption{Conservation of energy in LAMMPS running on LUMI-G.}
    \label{fig:lammpsenergy}
    \end{minipage}
\end{figure}

To test whether the integrator for each MD software is numerically stable on ROCm, extended simulations were run on LUMI-G, with the same parameters as the benchmarks, but with the 20k and 61k atom systems running for 2ns and the 465k, 1400k and 3000k systems running for 0.4ns. Figures \ref{fig:gromacsenergy}-\ref{fig:lammpsenergy} show the total energy of each system over the course of the simulation, with the x-axis normalised to the length of the simulation.\\


\subsection{ARCHER2 OpenMP and MPI Performance}\label{sec:archer2ompappendix}

\begin{figure}[H]
    \centering
    \begin{minipage}{0.45\textwidth}
    \centering
    \includegraphics[width=\textwidth]{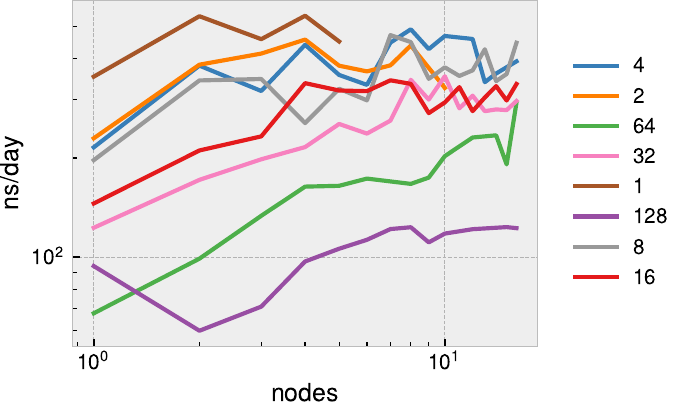}
    \caption{GROMACS performance with 20k atoms on ARCHER2. The legend is the number of threads.}
    \label{fig:archer2gromacs20kthreads}
    \end{minipage}\hfill
    \begin{minipage}{0.45\textwidth}
    \centering
    \includegraphics[width=\textwidth]{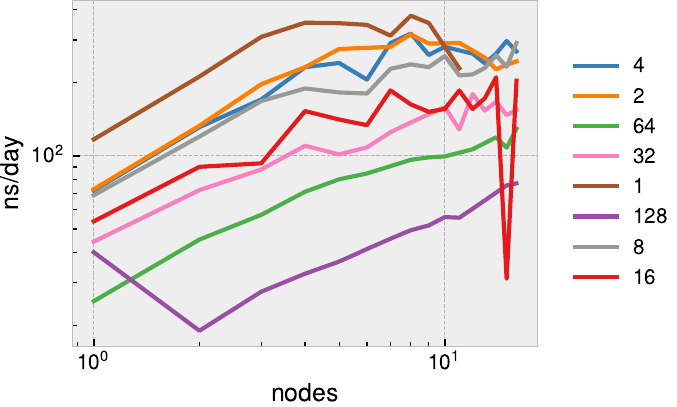}
    \caption{GROMACS performance with 61k atoms on ARCHER2. The legend is the number of threads.}
    \label{fig:archer2gromacs61kthreads}
    \end{minipage}
\end{figure}

\begin{figure}[H]
    \centering
    \begin{minipage}{0.45\textwidth}
    \centering
    \includegraphics[width=\textwidth]{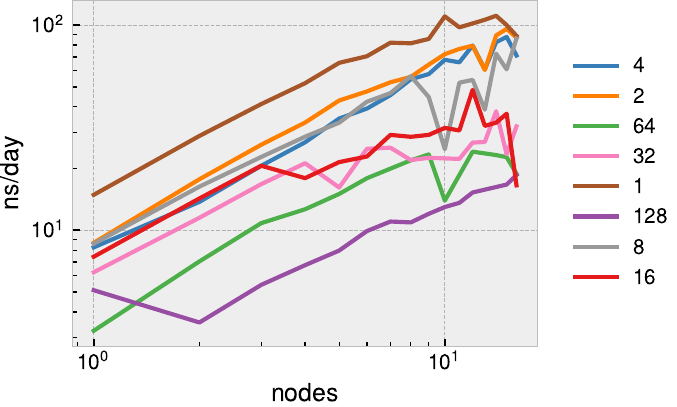}
    \caption{GROMACS performance with 465k atoms on ARCHER2. The legend is the number of threads.}
    \label{fig:archer2gromacs465kthreads}
    \end{minipage}\hfill
    \begin{minipage}{0.45\textwidth}
    \centering
    \includegraphics[width=\textwidth]{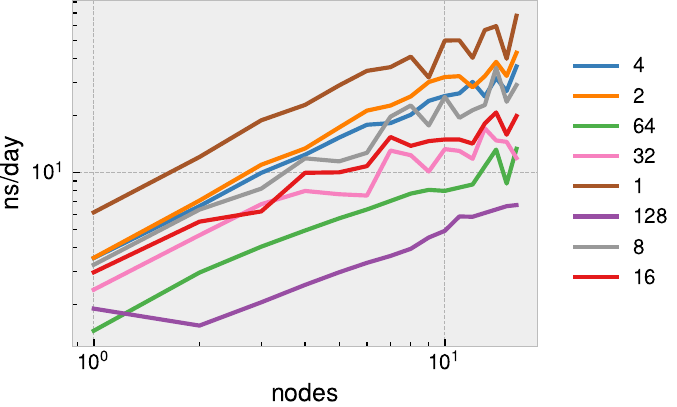}
    \caption{GROMACS performance with 1400k atoms on ARCHER2. The legend is the number of threads.}
    \label{fig:archer2gromacs1400kthreads}
    \end{minipage}
\end{figure}

\begin{figure}[H]
    \centering
    \begin{minipage}{0.45\textwidth}
    \centering
    \includegraphics[width=\textwidth]{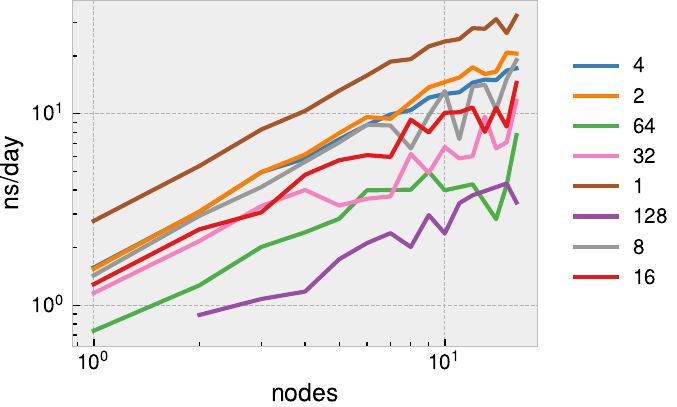}
    \caption{GROMACS performance with 3000k atoms on ARCHER2. The legend is the number of threads.}
    \label{fig:archer2gromacs3000kthreads}
    \end{minipage}\hfill
    \begin{minipage}{0.45\textwidth}
    \centering
    \includegraphics[width=\textwidth]{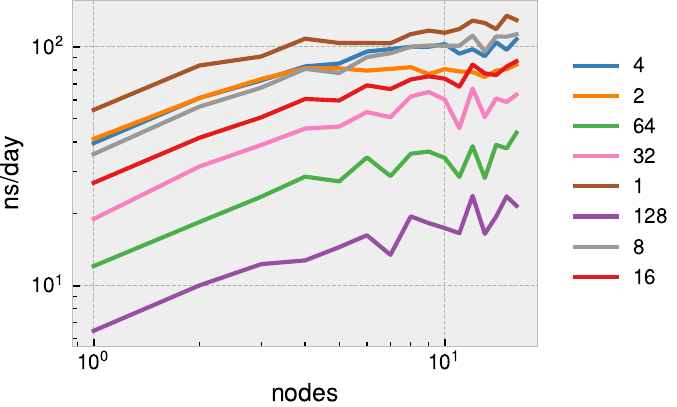}
    \caption{NAMD performance with 20k atoms on ARCHER2. The legend is the number of threads.}
    \label{fig:archer2namd20kthreads}
    \end{minipage}
\end{figure}

\begin{figure}[H]
    \centering
    \begin{minipage}{0.45\textwidth}
    \centering
    \includegraphics[width=\textwidth]{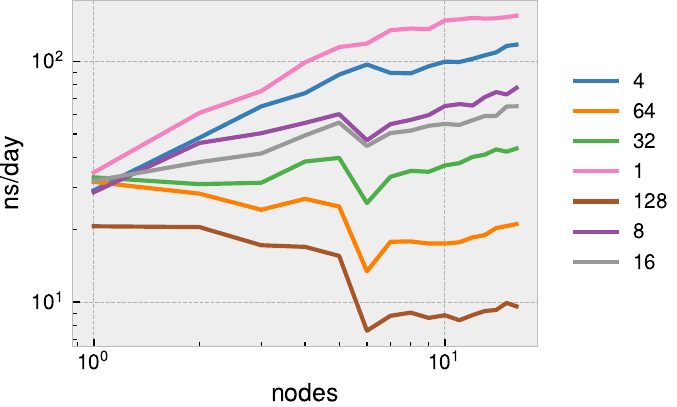}
    \caption{NAMD performance with 61k atoms on ARCHER2. The legend is the number of threads.}
    \label{fig:archer2namd61kthreads}
    \end{minipage}\hfill
    \begin{minipage}{0.45\textwidth}
    \centering
    \includegraphics[width=\textwidth]{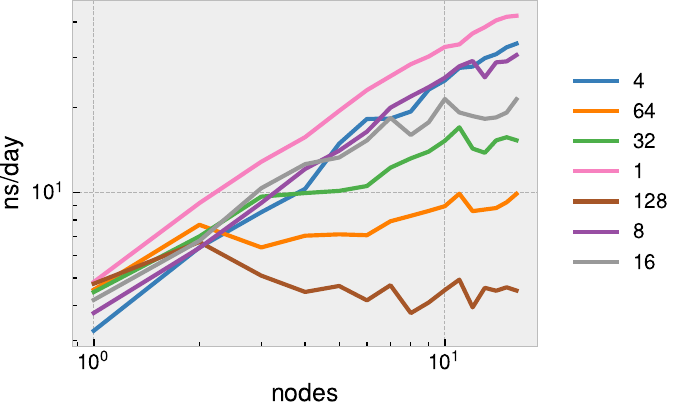}
    \caption{NAMD performance with 465k atoms on ARCHER2. The legend is the number of threads.}
    \label{fig:archer2namd465kthreads}
    \end{minipage}
\end{figure}

\begin{figure}[H]
    \centering
    \begin{minipage}{0.45\textwidth}
    \centering
    \includegraphics[width=\textwidth]{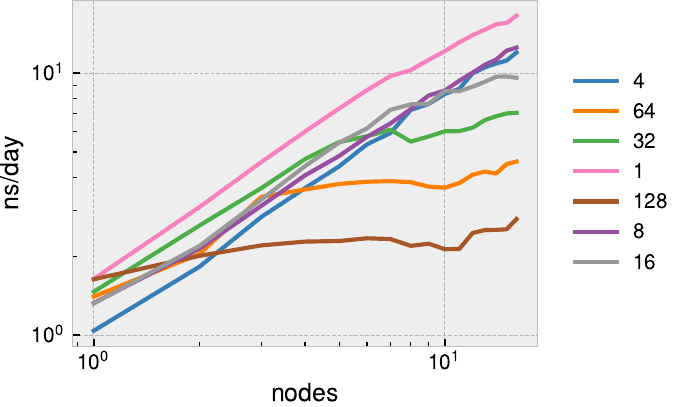}
    \caption{NAMD performance with 1400k atoms on ARCHER2. The legend is the number of threads.}
    \label{fig:archer2namd1400kthreads}
    \end{minipage}\hfill
    \begin{minipage}{0.45\textwidth}
    \centering
    \includegraphics[width=\textwidth]{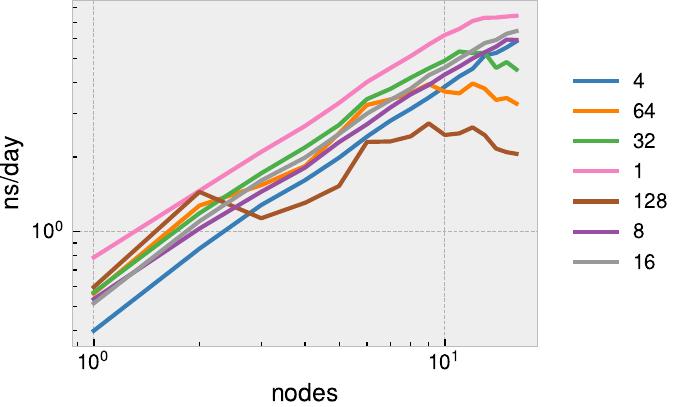}
    \caption{NAMD performance with 3000k atoms on ARCHER2. The legend is the number of threads.}
    \label{fig:archer2namd3000kthreads}
    \end{minipage}
\end{figure}

\begin{figure}[H]
    \centering
    \begin{minipage}{0.45\textwidth}
    \centering
    \includegraphics[width=\textwidth]{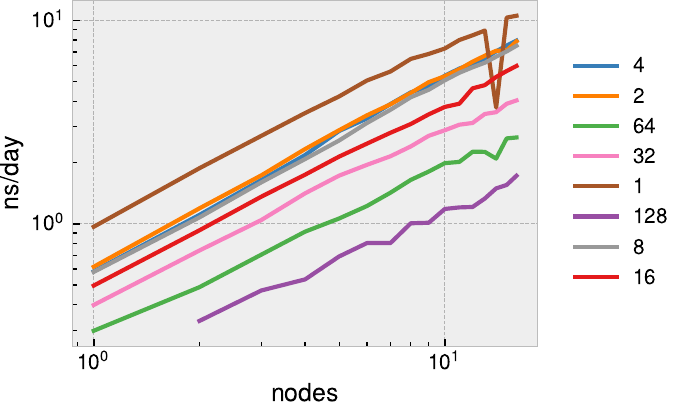}
    \caption{LAMMPS performance with 20k atoms on ARCHER2. The legend is the number of threads.}
    \label{fig:archer2lammps20kthreads}
    \end{minipage}\hfill
    \begin{minipage}{0.45\textwidth}
    \centering
    \includegraphics[width=\textwidth]{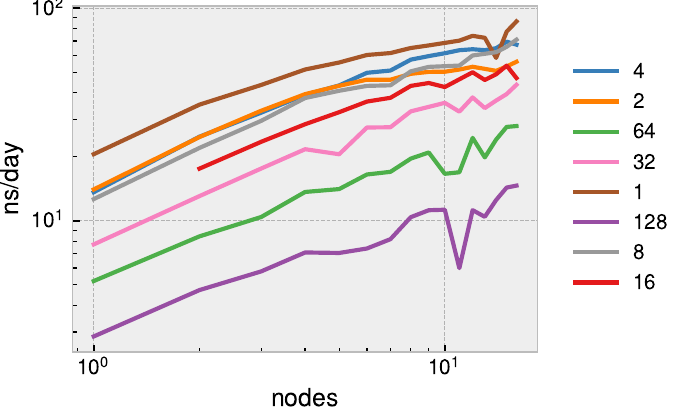}
    \caption{LAMMPS performance with 61k atoms on ARCHER2. The legend is the number of threads.}
    \label{fig:archer2lammps61kthreads}
    \end{minipage}
\end{figure}

\begin{figure}[H]
    \centering
    \begin{minipage}{0.45\textwidth}
    \centering
    \includegraphics[width=\textwidth]{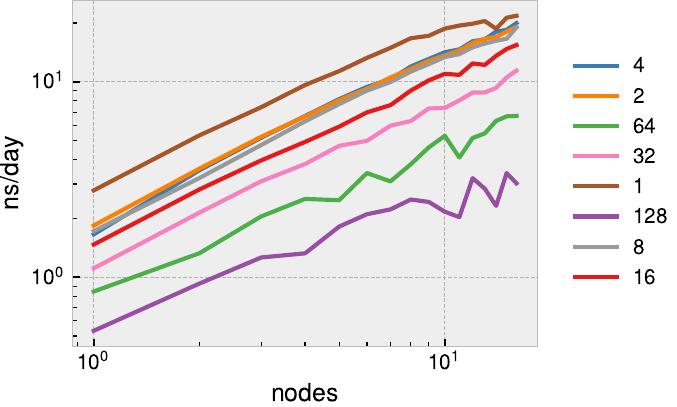}
    \caption{LAMMPS performance with 465k atoms on ARCHER2. The legend is the number of threads.}
    \label{fig:archer2lammps465kthreads}
    \end{minipage}\hfill
    \begin{minipage}{0.45\textwidth}
    \centering
    \includegraphics[width=\textwidth]{figs/lammps_optimal_threads_1400k}
    \caption{LAMMPS performance with 1400k atoms on ARCHER2. The legend is the number of threads.}
    \label{fig:archer2lammps1400kthreads}
    \end{minipage}
\end{figure}

\begin{figure}[H]
    \centering
    \begin{minipage}{0.45\textwidth}
    \centering
    \includegraphics[width=\textwidth]{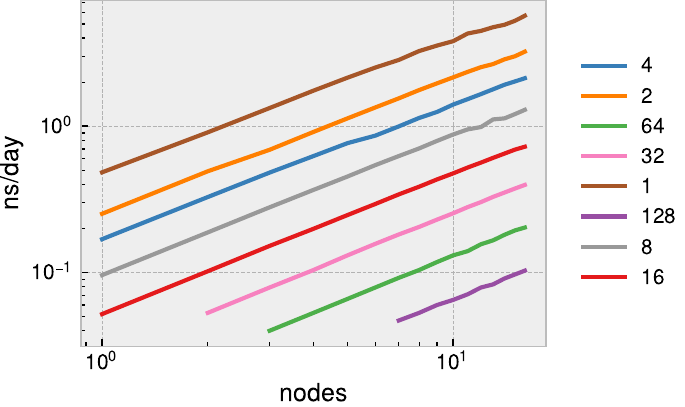}
    \caption{LAMMPS performance with 3000k atoms on ARCHER2. The legend is the number of threads.}
    \label{fig:archer2lammps3000kthreads}
    \end{minipage}\hfill
\end{figure}

GROMACS, NAMD and LAMMPS all support mixed OpenMP and MPI parallelisation. To determine the optimum combination of MPI processes and OpenMP threads, a simulation was run for each program, for each test system, using between 1-16 nodes and different combinations of OpenMP threads and MPI nodes (128 processes of 1 thread, 64 processes of 2 threads, 32 processes of 4 threads, etc) such that the entire node is used. Figures \ref{fig:archer2gromacs20kthreads}-\ref{fig:archer2lammps3000kthreads} show the results of these simulations.

\subsection{Jargon Buster}

\textbf{HPC} - High Performance Computing (supercomputing)\\
\textbf{FLOPS} - Floating point operations (e.g. maths operations) per second\\
\textbf{X86\_64 and ARM} - different CPU designs\\
\textbf{ns} - nanoseconds, a common representation of simulation time.\\
\textbf{MD} - molecular dynamics, the most popular method in computational biology.\\
\textbf{GPU} - graphics processing unit. The name is now mostly a misnomer, GPUs are commonly used in scientific software, as their architechture (which was designed to render graphs) allows for many small, fast calculations to be performed in parallel.\\
\textbf{CUDA and ROCm/HIP} - software layer that allows programs to run calculations on the GPU.\\
\textbf{GH200} - Nvidia's new AI accelerator card.\\
\textbf{LUMI-G} - EuroHPC supercomputer based in Finland.\\
\textbf{MPI and OpenMP} - software layers that allow calculations to run on many cores of one CPU or across multiple CPUs\\
\textbf{Scaling} - how well a simulation runs when spread across many CPU cores, GPUs or HPC nodes, or how well a fixed number of CPU cores/GPUs/nodes can run increasingly large simulations.\\
\textbf{VRAM} - video RAM (GPU memory)\\
\textbf{MIG} - multi-instance GPU, a feature of Nvidia GPUs that allows the GPU to be subdivided into smaller 'instances'\\
\textbf{CCSD} - coupled cluster singles and doubles (CCSD), a method used in computational chemistry\\
\textbf{EM} - electron microscopy\\
\textbf{API} - application protocol interface, an interface that allows computer programs to communicate with one another\\
\textbf{SLURM} - a program that manages user access to HPC systems\\

\typeout{get arXiv to do 4 passes: Label(s) may have changed. Rerun}

\end{document}